\documentclass[3p, times]{elsarticle}
\usepackage{amssymb}
\usepackage{amsmath}
\usepackage{graphicx}
\usepackage[retainorgcmds]{IEEEtrantools}
\usepackage[colorlinks]{hyperref}
\hypersetup{
  citecolor = {blue}
}
\usepackage{siunitx}
\usepackage{pdflscape}

\usepackage{enumitem}
\usepackage[normalem]{ulem}
\useunder{\uline}{\ul}{}
\usepackage{setspace} 
\usepackage[utf8]{inputenc}
\usepackage[linesnumbered, ruled, vlined]{algorithm2e}
\usepackage{rotating}
\usepackage{nicefrac}
\usepackage{xspace}
\usepackage{float}
\usepackage{caption}
\usepackage{subcaption}
\usepackage{rotating, lscape, longtable, booktabs, siunitx, multirow}
\usepackage[capitalise, noabbrev]{cleveref}
\usepackage{verbatim}
\usepackage{mathtools}

\usepackage{tkz-graph}

\usetikzlibrary{arrows,arrows.meta,shapes,decorations.pathmorphing, decorations.markings, positioning}
\let\svtikzpicture\tikzpicture
\def\tikzpicture{\noindent\svtikzpicture}

\newcommand{\uset}[1]{\ifmmode\left\{\,#1\,\right\}\else\{\,#1\,\}\fi}
\newcommand{\ulst}[1]{\ifmmode\left[\,#1\,\right]\else[\,#1\,]\fi}
\newcommand{\upar}[1]{\ifmmode\left(\,#1\,\right)\else(\,#1\,)\fi}
\newcommand{\uioc}[1]{\ifmmode\left(\,#1\,\right]\else(\,#1\,]\fi}
\newcommand{\uico}[1]{\ifmmode\left[\,#1\,\right)\else[\,#1\,)\fi}

\journal{Advanced Engineering Informatics}

\begin{document}

\begin{frontmatter}

\title{Robust Time Series Denoising with Learnable Wavelet Packet Transform}

\author[unicamp]{Ga\"{e}tan Frusque\corref{cor1}}
\ead{gaetan.frusque@epfl.ch}

\author[unicamp]{Olga Fink\corref{cor1}}
\ead{olga.fink@epfl.ch}

\address[unicamp]{Laboratory of Intelligent Maintenance and Operations Systems, EPFL, Lausanne}

\cortext[cor1]{I am corresponding author}



\begin{abstract}
Signal denoising is a key preprocessing step for many applications, as the performance of a learning task is closely related to the quality of the input data. 
In this paper, we apply a signal processing based deep neural network architecture, a learnable extension of the wavelet packet transform. 
As main advantages, this model has few parameters, an intuitive initialization and strong learning capabilities. 
Moreover, we show that it is possible to easily modify the parameters of the model after the training step to tailor to different noise intensities. 
Two case studies are conducted to compare this model with the state of the art and commonly used denoising procedures. 
The first experiment uses standard signals to study denoising properties of the algorithms. The second experiment is a real application with the objective to remove audio background noises. 
We show that the learnable wavelet packet transform has the learning capabilities of deep learning methods while maintaining the robustness of standard signal processing approaches. 
More specifically, we demonstrate that our approach maintains excellent denoising performances on signal classes separate from those used during the training step. 
Moreover, the learnable wavelet packet transform was found to be robust when different noise intensities, noise varieties and artifacts are considered.
\end{abstract}

\begin{keyword}
denoising \sep deep learning \sep spectrogram \sep time series \sep acoustic signals
\end{keyword}

\end{frontmatter}

\section{Introduction}

Real world signals are often corrupted by noise which needs be removed before any further analysis or processing step. To solve this problem, several contributions based on different approaches have been proposed. 
Several approaches for signal denoising have been proposed. This includes approaches such as dictionary learning \cite{li2009image}, empirical mode decomposition \cite{li2018data}, singular and higher order singular values decomposition \cite{jha2010denoising}, \cite{rajwade2012image} or canonical polyadic decomposition \cite{frusque2021canonical}.
Particularly wavelet-based methods are considered as an essential tool for multi-resolution and time-frequency analysis \cite{bayer2019iterative}. They often provide relevant features to monitor industrial systems with time signals \cite{gharesi2020neuro}, \cite{lv2021predictive}, or can bu used for data augmentation \cite{zhao2022highly}. The wavelet shrinkage operation that was theoretically investigated in \cite{donoho1994ideal} is still considered as one of the most powerful tools to perform signal denoising in many fields. Thus, wavelet packet transform (WPT) denoising has been used recently \cite{zhou2020partial}, \cite{kumar2021stationary} due to its ability to denoise regular frequency bands of desired size and to remove backgrounds with a specific frequency content.

To perform wavelet denoising, several hyperparameters need to be set. The parameters include the threshold used for wavelet shrinkage, the thresholding function and the wavelet family considered for the decomposition. Several heuristics have been proposed to address threshold selection \cite{alyasseri2019eeg}, such as the universal threshold \cite{donoho1994ideal}, Stein's unbiased risk estimation or the Bayesian shrink method \cite{chang2000adaptive}.
However, the selection of the wavelet family and the correct heuristics require specialized knowledge to remain robust to the complexity of the real data. Recent work opts for learning or automating the best hyperparameter configuration by supervised learning from a training dataset. In \cite{alyasseri2019eeg} a genetic algorithm is used to find the best wavelet denoising strategy for EEG denoising.
These methods benefit from the recent evolution of storage capacities and computing power allowing the constant increase of the amount of data collected to form the training dataset.  

For supervised denoising, deep learning methods have recently made a significant progress, particularly in application areas with large amounts of data, such as image denoising \cite{zhang2017beyond} or speech enhancement \cite{nossier2020comparative}. The use of deep neural networks (NNs) for denoising in domains where data is more specific and difficult to collect, such as biological signals, is a current issue \cite{xie2021bioacoustic}, \cite{arsene2019deep}. The main supervised denoising architectures based on deep learning include the convolutional neural network (CNN) \cite{kounovsky2017single}, the convolutional Auto-Encoder (AE) \cite{liu2019fault} and the U-Net \cite{nossier2020experimental}. However, these deep residual models contain a large number of parameters to be trained, and may lose efficiency when applied to an industrial dataset with a limited number of examples \cite{lyu2022novel}.  

Recently, deep architectures inspired by signal processing approaches have been proposed \cite{ravanelli2018speaker}. The main advantages of these approaches are to find more meaningful CNN filters, to  gain in interpretability and to reduce the number of parameters.   
In this work, we combine two of the main signal denoising methodologies, namely WPT denoising with wavelet shrinkage and deep autoencoder denoising.
Thus, we use a WPT-based deep learning architecture with learnable activation functions mimicking wavelet shrinkage, referred to as Learnable WPT (L-WPT). 
The used method is a relaxed version of the L-WPT architecture of \cite{frusque2022learnable} to improve learning capabilities. This is the first application of L-WPT for a denoising task.
The advantage of this architecture is threefold: 
a) It is based on a very powerful signal processing approach to obtain a time-frequency representation with optimal resolution. This provides our L-WPT algorithm considerable learning capabilities with only few parameters compared to standard deep learning methods \cite{michau2021fully}, \cite{frusque2022learnable}. 
b) The L-WPT contains only interpretable parameters that can be adapted manually if the operating conditions change.
c) We propose an intuitive initialization of the parameters to make the behavior of L-WPT as close as possible to the standard WPT.

As a second contribution, we demonstrate in this work how our L-WPT is related to the universality of signal processing methods and the learning capabilities of deep learning approaches. This highlights the advantage of combining signal processing and deep learning methods.
Specifically, we evaluate on the one hand how well the L-WPT can specialize and learn the particularity of the training dataset, and on the other hand how well it is able to generalize to  information and artifacts that are different to the one contained in the training data.

After presenting the related work in Section~\ref{Section-related}, we provide the necessary background on WPT in Section~\ref{Section2}. The L-WPT for signal denoising is introduced in Section~\ref{Section3}. A comparative study between the proposed L-WPT and several deep NNs is made using a standard model for signal denoising in Section~\ref{P3}. Finally, the performance of the L-WPT is highlighted in the real case of background removal in Section~\ref{Section4} before concluding.

\section{Related work}\label{Section-related}
\textbf{Wavelet denoising}: 
Wavelet shrinkage consists, after the application of a wavelet transform, in removing the low amplitude coefficients associated with noise. Different type of wavelet transform have been proposed. It can be any orthogonal wavelet transform \cite{bayer2019iterative}, or iterative methods such as the discrete wavelet transform (DWT).  The latter provides a representation of a signal in frequency bands of different temporal resolution. DWT has been applied among others for biomedical signals \cite{alyasseri2019eeg}, \cite{kumar2021stationary} and partial discharge applications \cite{zhou2020partial}. 
Unlike DWT, Wavelet Packet Transform (WPT) has the advantage of denoising on frequency bands of the same width. 
WPT denoising has been applied in various fields including speech enhancement \cite{oktar2016speech}, \cite{kumar2015comparative}, noise detection in bio-signals \cite{schimmack2016noise} and atomic force microscopy image denoising \cite{schimmack2018wavelet}. WPT has proven to be particularly suitable for background denoising with a specific frequency content. However the settings of the thresholds for each frequency band is a challenging task \cite{beale2020adaptive}, \cite{yue2019bayesian}. 

\textbf{Deep-learning-based supervised denoising}: The multilayer perceptron can be considered as the most basic deep neural network architecture. In the context of speech enhancement, this architecture has proven to be less robust and more difficult to train due to its high number of parameters compared to other methods such as the convolutional neural networks (CNNs) \cite{nossier2020experimental}, \cite{fu2017raw}.
The CNN uses the convolution operation. Only local and sparse connections between the input and output of each layers are considered. This reduces the number of parameters considerably and facilitates the learning process.  
CNN-based denoising methods have been used in several applications such as ECG denoising \cite{arsene2019deep}, speech enhencement \cite{kounovsky2017single} or image denoising \cite{zhang2017beyond}. In \cite{kong2019sound}, a CNN is used to separate sound events from the background in the time-frequency domain to improve sound classification.
The auto-encoder version of the CNN, the convolutional auto-encoder (AE) has been one the most widely used architectures for denoising. AE aims to produce an input-like representation using an encoder and a decoder. 
The objective of the encoder is to find an embedding of the input data that contains the important information, eliminating noise. A noise-free signal based on the embedding is estimated with the decoder \cite{kachuee2018dynamic}. 
It was used for speech recognition in \cite{grozdic2017whispered}. Several examples of AE-based denoising methods was successfully applied in the context of  prognostics and health Management \cite{fink2020potential}, \cite{lv2022vibration}.
In \cite{liu2019fault}, an autoencoder is used to denoise the vibration signals from the bearing dataset. The denoising step helps in monitoring the condition of the bearing by improving fault diagnosis. In \cite{berghout2020aircraft}, a denoising AE is used to improve the prediction of the remaining useful life of aircraft engines and in \cite{fan2020defective} is was used for defective wafer detection for semiconductor manufacturing process. 
One challenge encountered by AE is the potential loss of important information with the increasing number of layers. Adding skip connections to create a U-Net architecture \cite{ronneberger2015u} can potentially tackle this problem.
As shown in the comparative study in \cite{nossier2020experimental}, the U-Net architecture as proposed in \cite{pandey2019new} has the best overall performance compared to several other architectures without skipped connections. Several recent works on image denoising also demonstrate a good performance of U-Net denoising \cite{zhang2022practical}. 

\textbf{Deep NNs architectures inspired by signal processing}: Several approaches that include signal processing elements in NNs have already been proposed. In \cite{ravanelli2018speaker}, a CNN layer with kernels constrained to match bandpass filters only are used for speaker recognition. 
In \cite{hao2019multi} and \cite{xiong2016ecg}, wavelet transform and NNs are combined for ECG signal enhancement.
The first learnable extensions of WPT are focused on learning the best filter to use in through the entire architecture, such as in \cite{xiong2020novel}, \cite{recoskie2018learning}, \cite{jawali2019learning}, \cite{ha2021adaptive}. 
The first model to generalize filtering learning to each layer was in the context of the Discrete Wavelet Transform  (DWT) \cite{wang2018multilevel}. The authors in \cite{michau2021fully} proposed the DeSpaWN model which in addition add activation functions that perform automatic denoising. DeSpaWN has been successfully applied to classification and anomaly detection tasks for audio signals. An extension of the DeSpaWN architecture to WPT, called L-WPT, was proposed in \cite{frusque2022learnable}. The L-WPT shows a better performance compared to DeSpaWN for the same anomaly detection task. 

\section{Background}\label{Section2}
\subsection{Wavelet Packet Transform (WPT)}
The discrete WPT, introduced in \cite{coifman1992wavelet}, projects the signal on uniform frequency bands of desired size. The WPT has a multi-level structure and can be considered as a multi-resolution analysis since the output of the current level is recursively used as input to the next level. The basic bloc of a WPT applied to an input signal $\mathbf{y}$ is:
\begin{align}\label{eq:filt}
& (\mathbf{y}^{\rm lp})_{(n)} = ( \mathbf{h}^{lp} * \mathbf{y} )_{(2n)}, \hspace{0.3cm}  (\mathbf{y}^{hp})_{(n)} = ( \mathbf{h}^{\rm hp} * \mathbf{y} )_{(2n)}, 
\end{align}
where $*$ is the convolution operation and $\mathbf{y}^{\rm lp} $ ( or $\mathbf{y}^{\rm hp} $) corresponds to the low- (or high-) pass filtered input data with a cut-off frequency of $\pi /2$ followed by a sub-sampling by two. 
This transformation doubles the frequency resolution (the frequency content of each wavelet coefficient spans half the input data frequency) to the detriment of a halved time resolution ($\mathbf{y}^{\rm lp} $ and $\mathbf{y}^{\rm hp} $ each contains half the number of samples in $\mathbf{y}$). 
By applying the same block Eq.~(\ref{eq:filt}) on  $\mathbf{y}^{\rm lp} $ and  $\mathbf{y}^{\rm hp} $, we then obtain four outputs that divide the frequency content of the input signal in four even bands. 
The underlying algorithm behind of WPT has a tree structure characterised by $L$ layers, corresponding to the number of times we apply the block Eq.~(\ref{eq:filt}) to the outputs of the previous layer. We refer to the nodes as the succession of a filtering and a sub-sampling operation. The outputs from the $2^L$ nodes at layer $L$ form the time-frequency representation of our signal $\mathbf{y}$.

A perfect reconstruction of the WPT of a signal from any layer $L$ is possible. This operation is called inverse WPT (iWPT) and is possible only if the filters $\mathbf{y}^{\rm lp} $, $\mathbf{y}^{\rm hp} $ and the transposed filters follow the conjugate mirror conditions \cite{mallat1999wavelet}, \cite{strang1996wavelets}.


\subsection{Denoising with WPT}\label{p2C}
Signal denoising is one of the major applications of wavelet analysis \cite{bayer2019iterative}. It has been shown that, a wavelet transform will lead to a sparse decomposition of regular and structured signals \cite{mallat1999wavelet}. We can then assume that the noise will correspond to wavelet coefficients of small amplitudes. 
Several procedures eliminating small coefficients of a WPT already exists \cite{bayer2019iterative}, \cite{donoho1994ideal}. The two most commonly applied approaches use the soft- and hard-thresholding operators \cite{donoho1994ideal}. The soft-thresholding appears to be more adequate for image denoising with small signal-to-noise ratios (SNR). We propose in this paper to study only the hard thresholding (HT) operation to eliminate the low coefficients of the WPT. Considering a WPT with $L$ layers, and $y^i_L(t)$ one of the obtained coefficients at node $i$, the HT operation with threshold value $\lambda$ corresponds to: 
\begin{align}
HT_{\lambda}(y^i_L(t) ) = \begin{cases} y^i_L(t) & \text{if $|y^i_L(t)  | > \lambda$,} \\
                      0 & \text{else} \end{cases}
\end{align}
The operation has to be performed for all coefficients with index $t$, and all nodes $i$ of the layer $L$. An estimation of the denoised input signal can then be computed by applying the iWPT to the thresholded coefficients.

\section{Learnable Wavelet Packet Transform}\label{Section3}
\subsection{Adapted denoising}\label{p2F}
One of the major problems of the HT denoising method presented in Section~\ref{p2C} is that it does not adapt to the frequency of the input signal. This can be problematic if the background noise we want to remove from the pure signal has a structured frequency content. Some methodologies adapting the thresholding value according to the frequency have been previously proposed \cite{beale2020adaptive}, \cite{yue2019bayesian}, \cite{alyasseri2019eeg}. 
We propose to go further by applying an adapted activation function with learnable biases to each node of the entire tree structure of the WPT algorithm. The proposed activation function will perform a learnable thresholding eliminating the coefficients related to noise. 

In order to have a differentiable thresholding function, we propose the double sharp sigmoid activation function proposed in \cite{michau2021fully} and denoted as ${\eta}_{\gamma}(x)$:
\begin{align}\label{eq:act}
{\eta}_{\gamma}(x)=x \left[  \frac{1}{1+e^{10(x+\gamma)}} +  \frac{1}{1+e^{-10(x-\gamma)}}   \right],
\end{align}
with $\gamma$ the learned bias acting as a threshold on both sides of the origin.
\subsection{L-WPT: an autoencoder model inspired by WPT}
The proposed L-WPT methodology is an instance of autoencoders, where the encoding part mimics the tree structure algorithm of the WPT and the decoding part mimics the inverse tree structure of the iWPT. 
Since the WPT will provide a sparse representation only for structured signals respecting specific properties \cite{mallat1999wavelet} which does not always hold for real applications, we propose to learn the WPT filters. The idea is to find an adapted sparse representation of the signal of interest, which, combined with the proposed denoising activation function Eq.~\ref{eq:act}, will be able to better convert a potentially complex noise into low coefficients that can be then eliminated.

\begin{figure}[h]
\centering
\includegraphics[width=0.40\textwidth]{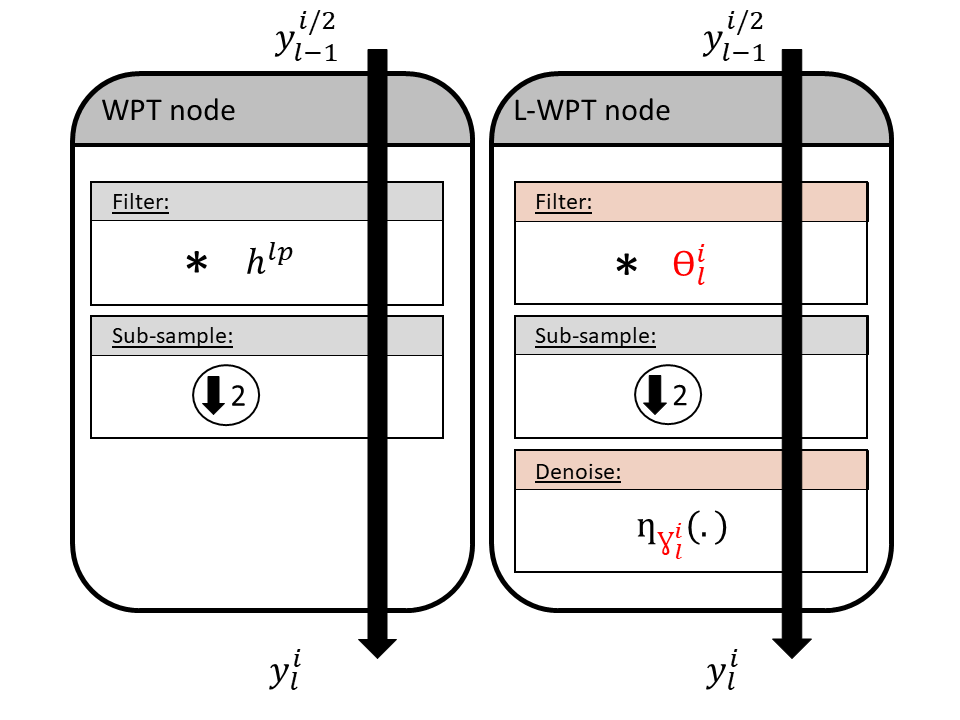}
\caption{Comparison between the operations performed in the WPT node (left) and a L-WPT node (right). The variables in red are learnable. If we consider a WPT node using the  $\mathbf{h}^{\rm lp}$ filter the input signal becomes $\mathbf{y}_{l-1}^{(i-1)/2}$.}\label{fig-node}
\end{figure}

Thus, considering the encoding part, the filter used in each node is replaced by a convolutional layer with the stride two followed by the denoising activation function Eq.~\ref{eq:act}. To obtain the coefficients at layer $l$ and node $i$ (denoted $\mathbf{y}_l^{i}$), we need to convolve the coefficients at the previous layer with the kernel of the current node, denoted by $\theta^{i}_{l}$, and apply the activation function Eq.~\ref{eq:act}. It can be written as:
\begin{align}
\mathbf{y}_l^{i} = \eta_{\gamma_l^i}\left( \theta^{i}_{l} * \mathbf{y}_{l-1}^{\lfloor  \frac{i}{2} \rfloor }  \right),
\end{align}
where $\lfloor \bullet \rfloor$ corresponds to floor function. The activation function is applied at each coefficient of $\mathbf{y}_l^{i}$ with the learnable bias value $\gamma_l^i$. An illustration providing an comparison of the operations performed in a WPT node and a L-WPT node is shown in Figure~\ref{fig-node}

For the decoding part, we only replace the filters by a transposed convolutional layer with stride 2. It is possible to compute a denoised estimation of the coefficients at layer $l$ and node $i$ (denoted $\hat{\mathbf{y}}_l^{i}$) by using the denoised estimation of the higher layers and two kernels denoted as $\beta^{2i}_{l+1}$ and $\beta^{2i+1}_{l+1}$. It can formulated be  written as:
\begin{align}
\hat{\mathbf{y}}_l^{i} = \mathbf{\beta}^{2i}_{l+1} *   {\rm up} \hspace{-0.1cm} \left[\hat{\mathbf{y}}^{2i}_{l+1} \right] +  \mathbf{\beta}^{2i+1}_{l+1} *   {\rm up} \hspace{-0.1cm} \left[\hat{\mathbf{y}}^{2i}_{l+1}\right] 
\end{align}
where the up-sampling operation ${\rm up} [\bullet ]$ is necessary to counteract the effect of the down-sampling in Eq.~(\ref{eq:filt}). It can be defined as ${\rm up} [\mathbf{x} ]_{(2n)}=x_{(n)}$ and  ${\rm up} [\mathbf{x} ]_{(2n+1)}=0$.

Considering these notations, the input signal can be denoted as $\mathbf{y}_0^0$, the output signal as $\hat{\mathbf{y}}_0^0$ and we have $\mathbf{y}_L^i=\hat{\mathbf{y}}_L^i$ $\forall i \in \{0,...,2^{L}-1 \}$.


\subsection{Adaptable weights}\label{pDELTA}
Since our architecture is inspired by a signal processing methodology, we have a good understanding of the purpose of each part of the network, and how a modification of the parameters impacts the output signal. The bias of the activation function is meaningful and is used to threshold low amplitude coefficients associated with noise. After learning the kernels and bias, the L-WPT can be applied to denoise signals under different operating conditions and noise levels. Thus, it is always possible to modify the biases afterwards if, for example, the noise level increases or decreases. 

We propose  a simple modification of the biases, called $\delta$ modification. A trained L-WPT with the $\delta$ modification is then denoted as L-WPT-($\delta$). This modification consists simply in multiplying each bias value by $\delta$ and can be written as:
\begin{align}
 \gamma^i_l \leftarrow \delta \gamma^i_l \hspace{1cm} & \forall l \in \{1,...,L\} \\
 &\forall i \in \{0,...,2^L-1\} \nonumber
\end{align} 
$\delta$ is chosen according to the variations of the noise level with respect to the training data (see backgrounds suppression application section~\ref{ss3}).

\subsection{Intuitive weight initialisation}\label{pINI}
One of the advantages of the proposed method is that there are intuitive initializations of filters and biases in order to start with a L-WPT that behaves in a similar way as a standard WPT. Considering a kernel of size $K+1$ with $K$ an odd number and $ k \in \{0,...,K\}$, denoted as $\mathbf{h}^{\rm PR} $, which satisfies the conjugate mirror property (examples of such kernels inlcude wavelet families like Daubechies, Haar or Coiflets), the initialisation of the kernels in the encoding and decoding parts for all layers $l$ and nodes $i$ is \cite{strang1996wavelets}, \cite{mallat1999wavelet}:
\begin{align}
& \theta_l^i(k)=\mathbf{h}^{\rm PR}(k) \hspace{2.7cm} \text{if $i$ is even,} \\
& \theta_l^i(k)=  (-1)^k \mathbf{h}^{\rm PR}(K-k)    \hspace{1.3cm} \text{if $i$ is odd,} \\
& \beta_l^i(k)= \mathbf{h}^{\rm PR}(K-k)  \hspace{2.05cm} \text{if $i$ is even,} \\
& \beta_l^i(k)= (-1)^{k+1} \mathbf{h}^{\rm PR}(K-k)   \hspace{1cm} \text{if $i$ is odd.}
\end{align}

Finally, the denoising activation function has to be replaced by a linear function, which can be done if we initialise all biases $\gamma_l^i$ with 0.

\subsection{Objective function and training}\label{p2G}
We denote by $s$ a pure signal, and $\tilde{s}=s+b$ the same signal corrupted by a background noise $b$. Thus, assuming that we have a set of pure signals and background noise, we are looking for the best kernels of the encoding part ${ \mathbf{\Theta}=[\mathbf{\theta}_0^0, \mathbf{\theta}_1^0, \mathbf{\theta}_1^1,...,  \mathbf{\theta}_{L}^{2^{L}-1}]}$, kernels of the decoding part ${ \mathbf{\textit{B}}=[\mathbf{\beta}_0^0, \mathbf{\beta}_1^0, \mathbf{\beta}_1^1,...,  \mathbf{\beta}_{L}^{2^{L}-1}]}$ and biases ${\mathbf{\Gamma}   = [ \gamma^{0}_{1}, \gamma^{1}_{1},...,  \gamma^{2^{L}-1}_{L} ]}$ minimising the following loss function:
\begin{align}\label{Final_func}
\underset{ \mathbf{\Theta},{\mathbf{\textit{B}}}, \mathbf{\Gamma}    }{\rm argmin} \hspace{0.5cm}   \sum_{n}^{} {\mid\mid \hat{s}_n - s_n  \mid\mid_F^2, }
\end{align}
where $\hat{s}=\hat{\mathbf{y}}_0^0$ is the reconstructed signal from the input data $\tilde{s}=\mathbf{y}_0^0$.

We use the Adam optimiser \cite{kingma2014adam} with a learning rate of 0.0005 and a batch size of 8 to train the L-WPT. The number of epochs is set to 500 and the learning rate is divided by 10 after the epoch 350 and the epoch 450 for a better convergence. We initialise the filters and bias as presented in Section~\ref{pINI} using for $\mathbf{h}^{PR}$ Daubechies wavelet with 8 coefficients (called db4). 

By referring to $n_p$ as the number of trainable parameters and considering $L$ layers and $K+1$ coefficients per filters, we have for the L-WPT $n_p=\sum_{l=1}^{L} {2^{L}K} + \sum_{l=1}^{L} {2^{L}} $ trainable parameters, where the first part are the filter parameters and the second the biases parameters.

\section{Denoising performances on standard functions}\label{P3}

We compare the L-WPT performance to other deep NNs architectures and wavelet shrinkage for denoising purpose. More details about these methods are given in sub-section~\ref{pComp}.  
For a benchmark study, we consider standard function classes commonly applied in denoising literature to evaluate the performance of denoising algorithms \cite{donoho1994ideal}, \cite{hao2017improved}, \cite{li2018data}. They mimic spatially variable functions arising in imaging, spectroscopy and other applications and are presented in sub-section~\ref{p3B}.
We quantify in sub-section~\ref{p4C} how the L-WPT leads to improve the denoising of signals from the training class, but also signals of different nature from other classes. 
It will demonstrate how L-WPT relates to the learning capabilities of deep learning approaches if it outperforms the standard WPT denoising on the training class. 
As well, it will show how the L-WPT relates to the universality of signal processing if it follows better generalisation than deep NNs. 
Robustness analysis of our method is extended by considering different noise levels in sub-section~\ref{p5C}.

\subsection{Comparison to other approaches}\label{pComp}
We compare our method with a signal processing approach and several deep NNs.  This will allow us to position our method with respect to the learning capabilities of deep models and the robustness of signal processing approaches.

We compare our framework to the hard thresholding wavelet shrinkage presented Section~\ref{p2C}. We call this method "Baseline-HT". We consider the deep NNs presented is Section~\ref{Section-related}: a standard CNN, a convolutional AE based on \cite{liu2019fault} and a U-Net model based on \cite{jimenez2019u}.
We provide in Appendix~\ref{AppendixA} a methodology to select the best AE architectures based on a set of pre-selected ones. 
In this case, the number of trainable parameters is extensive ($n_p=554954$) compared to our L-WPT with 5 layers ($n_p=1054$). To ensure that the results obtained are not mainly due to the difference in the number of parameters, causing the AE to overfit compared to the L-WPT, we also consider a similar AE architecture with the same number of parameters as the L-WPT. We refer to those two architectures as "AE-large" and "AE-small".

We derive the architectures of the U-Net and CNN models from the two obtained AE architectures. We refer to them as "U-Net-large", "U-Net-small", "CNN-large" and "CNN-small". The deep NNs are trained using exactly the same objective function optimisation parameters as the L-WPT (see Section~\ref{p2G}).

An overview of the key parameters of the six different architectures are provided in Appendix~\ref{AppendixC}

\subsection{Model functions and noise}\label{p3B}
We use the following benchmark case functions \cite{donoho1994ideal}, which are named \textit{Block}, \textit{ Bumps}, \textit{HeaviSine} and \textit{Doppler}. We propose to randomly generate signal classes $s$ inspired by those four function cases. The number of samples in each signal $s$ is set to $T=2^{13}$. More details about the generation of these function are given in Appendix~\ref{AppendixD}.

\begin{figure}[h]
\centering
\includegraphics[width=0.15\textwidth]{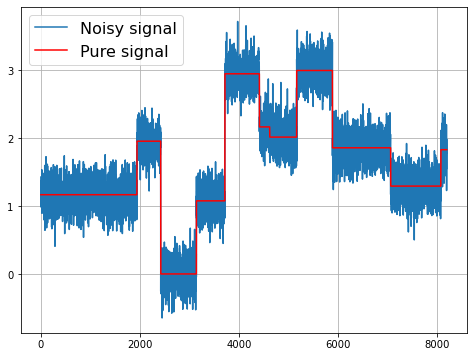}
\includegraphics[width=0.15\textwidth]{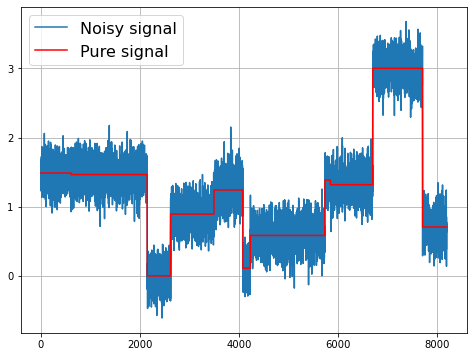}
\includegraphics[width=0.15\textwidth]{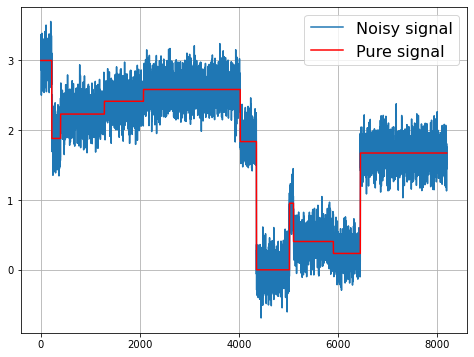}

\includegraphics[width=0.15\textwidth]{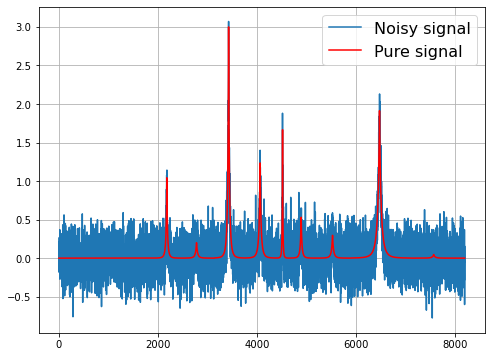}
\includegraphics[width=0.15\textwidth]{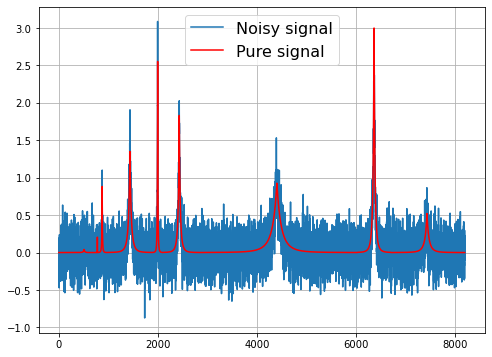}
\includegraphics[width=0.15\textwidth]{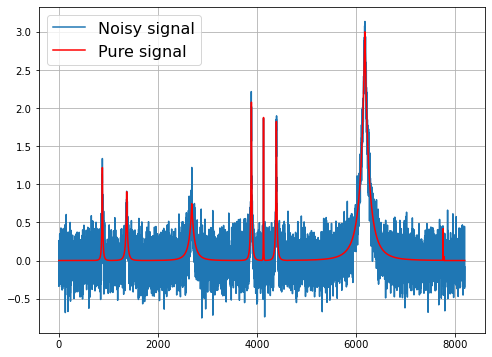}

\includegraphics[width=0.15\textwidth]{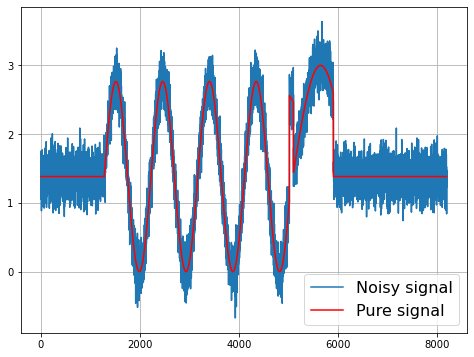}
\includegraphics[width=0.15\textwidth]{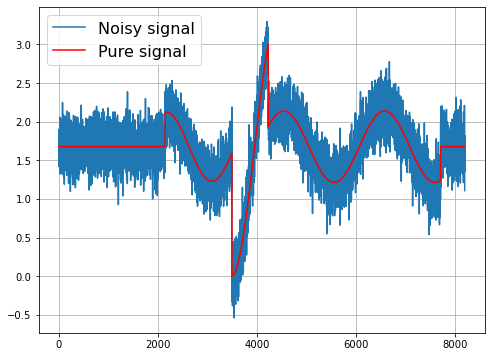}
\includegraphics[width=0.15\textwidth]{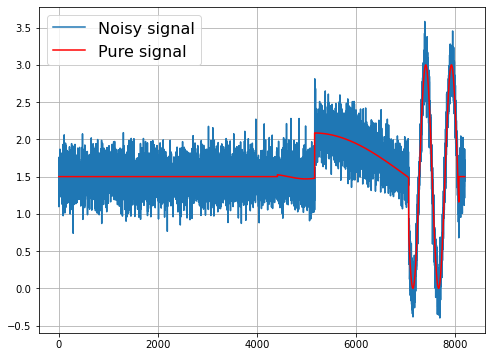}

\includegraphics[width=0.15\textwidth]{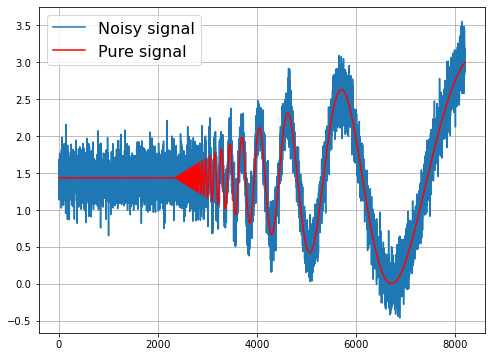}
\includegraphics[width=0.15\textwidth]{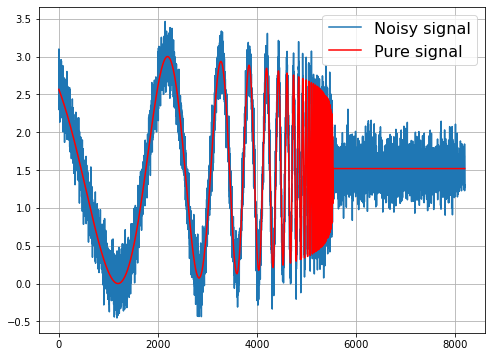}
\includegraphics[width=0.15\textwidth]{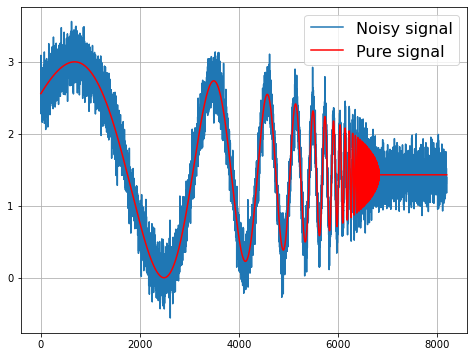}
\caption{Different realisations of the four function classes and their corrupted versions.}\label{fig-3A1}
\end{figure}

The pure signal are corrupted by adding a white Gaussian noise, with $\tilde{s}$ the corrupted signal. The corruption is performed as follows:
\begin{align}
\tilde{s}(t)=3s(t) + \sigma b(t)
\end{align}
where $ b(t)$ is a realisation of a normal distribution, and $\sigma$ is the noise level. The factor three is chosen to have an easier interpretation of the noise level, i.e., if $\sigma=1$ almost all realizations will be in the same amplitude range as the pure signal ($\approx$ 99.7\% chances to be below 3). In Figure~\ref{fig-3A1}, we display different realisations of pure signals and their corrupted counterparts with $\sigma=0.2$. 

\subsection{Robust denoising}\label{p4C}
We compare the denoising properties of the L-WPT and the methods presented in this Section~\ref{pComp}. Each method are trained on each class separately.  The training was done on 16000 realisations $\tilde{s}$ from one class with noise level $\sigma=0.2$. For the L-WPT, we set the number of layers to $L=5$.
We refer to  the class used for the training as $\mathbf{C}_T$.  
Then $N_{\rm test}=500$ test signals are generated for each class. 
We propose to evaluate the performance with respect to three scores: 1) the specialisation score denoted as $S_p$ which shows how efficient  a method is to denoise signals from $\mathbf{C}_T$; 2) $S_r$ the robustness score to demonstrate how the denoising performances are generalisable to the other classes; and 3)the mean score over all test signals, noted $\bar{S}=(S_p + 3S_r)/4$ that captures a trade-off between good specialisation and robustness. With $\hat{s}$ the estimation of the pure signal $s$, the computation of $S_p$ and  $S_r$ is derived as follows: 
\begin{align}
S_p = \frac{10^5}{T N_{\rm test}} \sum_{i / s_i \in \mathbf{C}_T}^{N} { || \hat{s}_i - s_i ||_F^2  } \\
S_r = \frac{10^5}{3 T N_{\rm test}} \sum_{i / s_i \notin \mathbf{C}_T}^{N} { || \hat{s}_i - s_i ||_F^2  }
\end{align}

In Table~\ref{TableC}, the $S_p$, $S_r$ and $\bar{S}$ scores for the cases when each model is trained using the four classes separately are displayed. The AE-large model has the best specialization performance
However, the method does not generalise well to the other classes. This is reflected in its $S_r$ score for the Bumps and Doppler classes. Overall, the L-WPT has the best robustness score, even better than the Baseline-HT method that is particularly adapted for Gaussian denoising. We can state that, for this experiment, the L-WPT keeps the robustness of a general, non trainable denoising procedure like the Baseline-HT, but also learns a relevant denoising for the signals of the learned class.

\begin{figure*}
\centering

\includegraphics[width=0.24\textwidth]{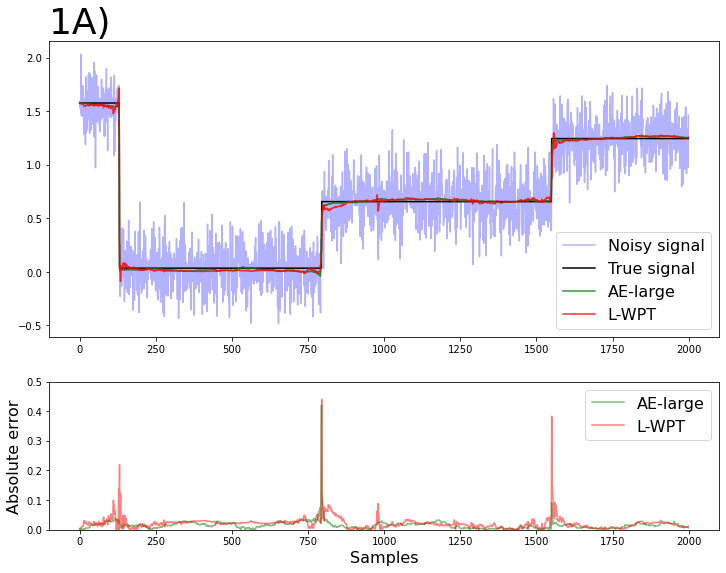}
\includegraphics[width=0.24\textwidth]{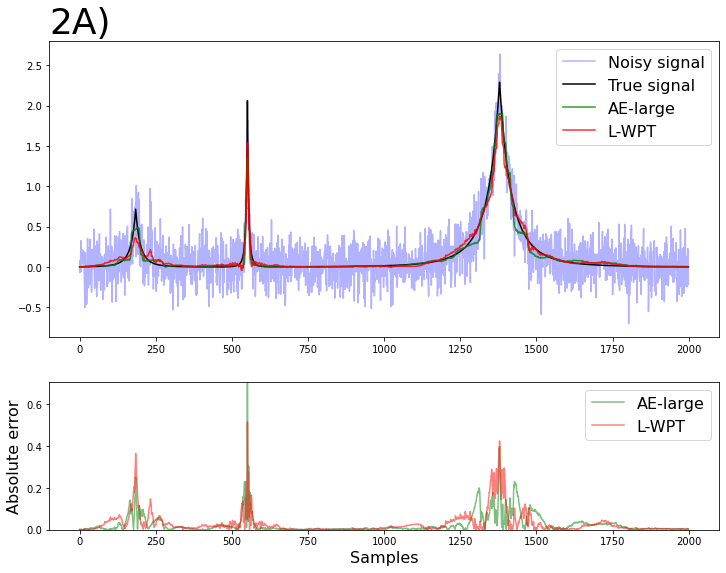}
\includegraphics[width=0.24\textwidth]{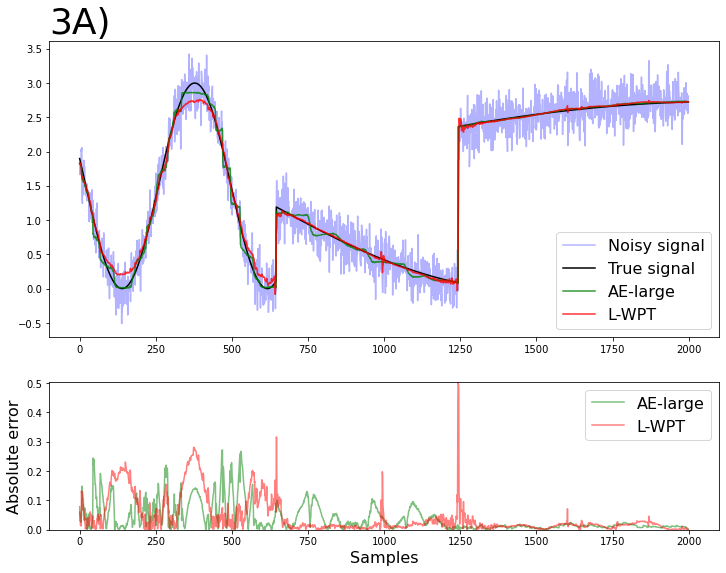}
\includegraphics[width=0.24\textwidth]{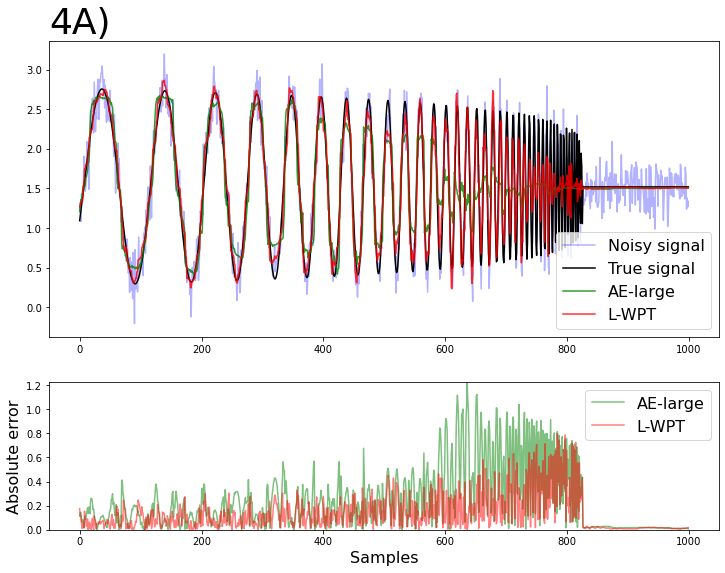}

\includegraphics[width=0.24\textwidth]{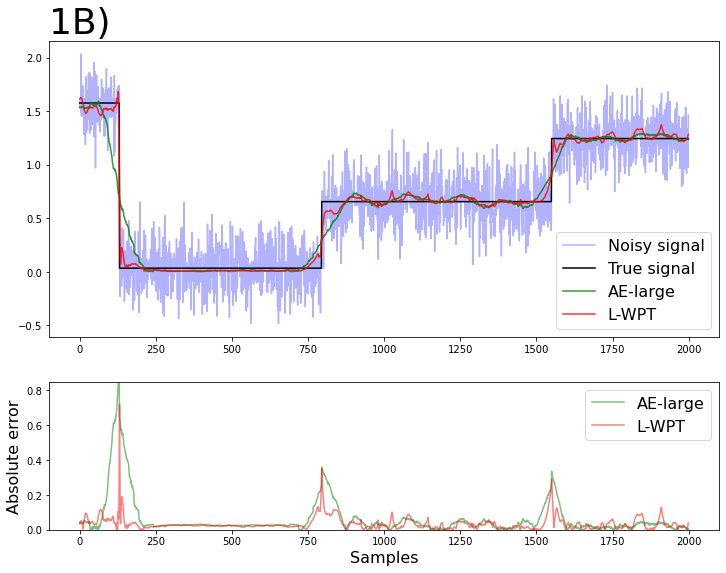}
\includegraphics[width=0.24\textwidth]{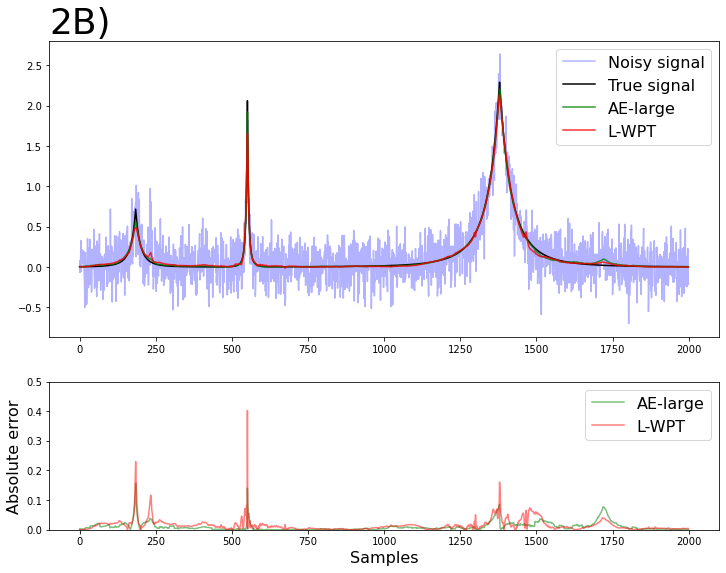}
\includegraphics[width=0.24\textwidth]{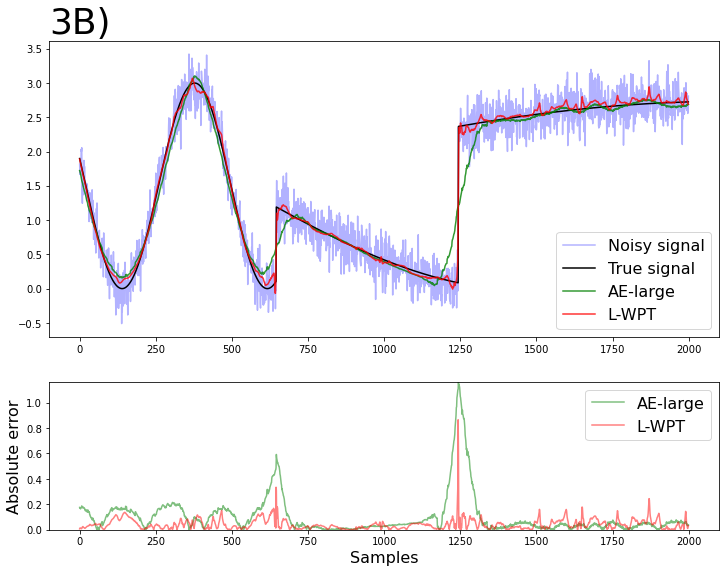}
\includegraphics[width=0.24\textwidth]{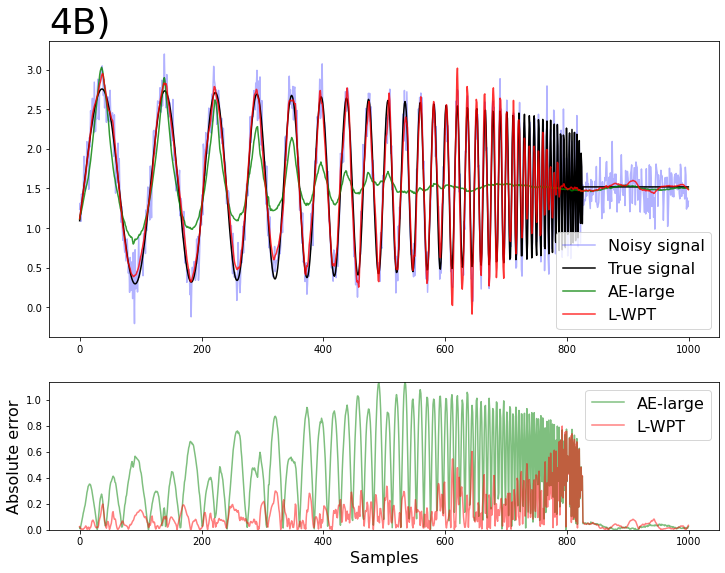}

\includegraphics[width=0.24\textwidth]{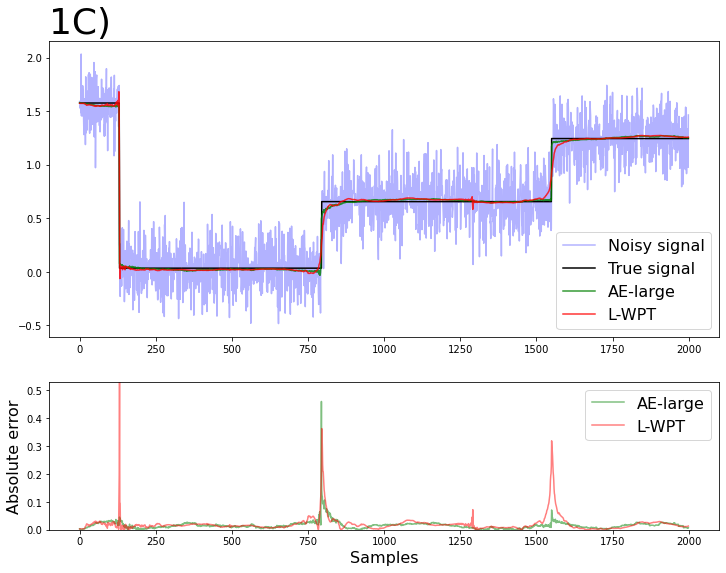}
\includegraphics[width=0.24\textwidth]{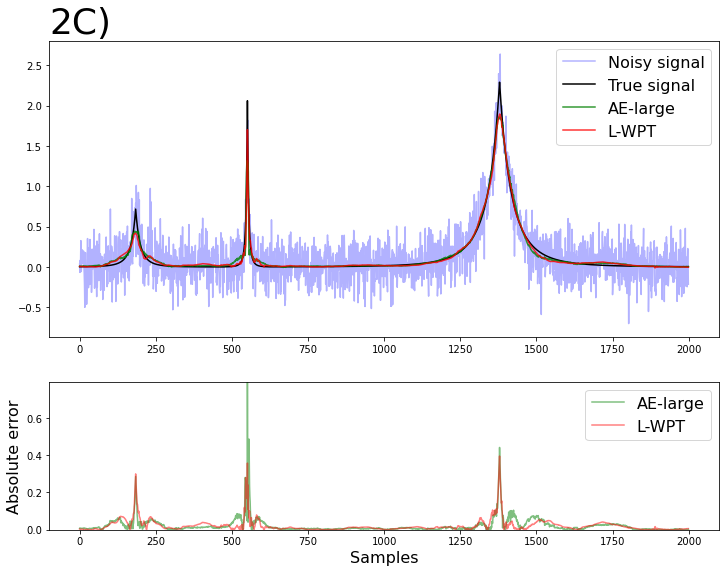}
\includegraphics[width=0.24\textwidth]{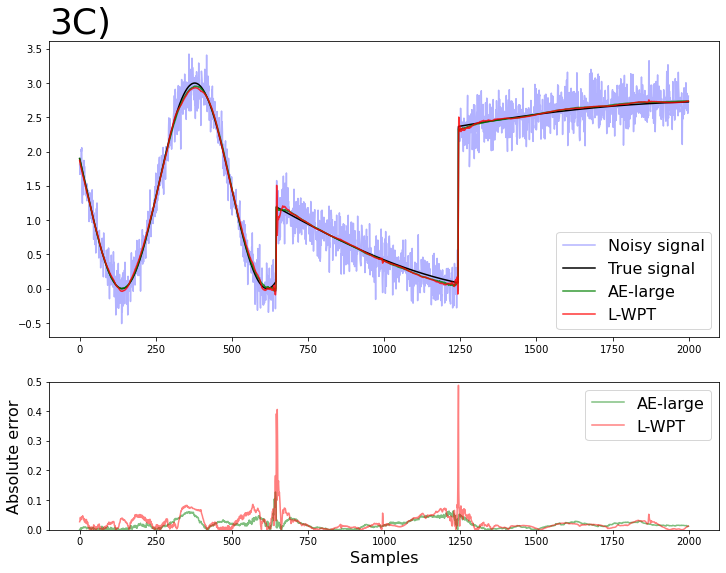}
\includegraphics[width=0.24\textwidth]{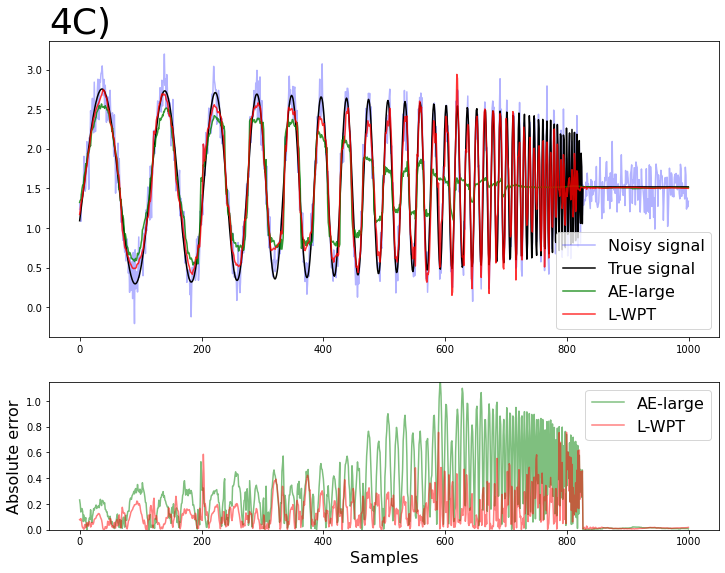}

\includegraphics[width=0.24\textwidth]{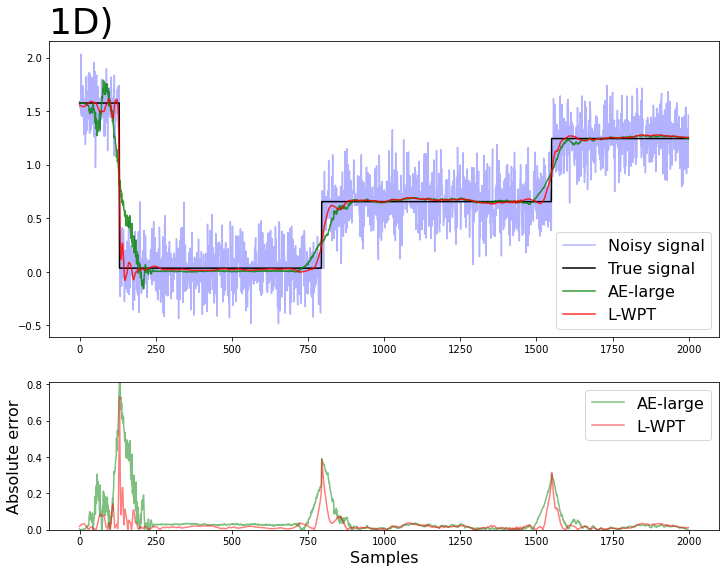}
\includegraphics[width=0.24\textwidth]{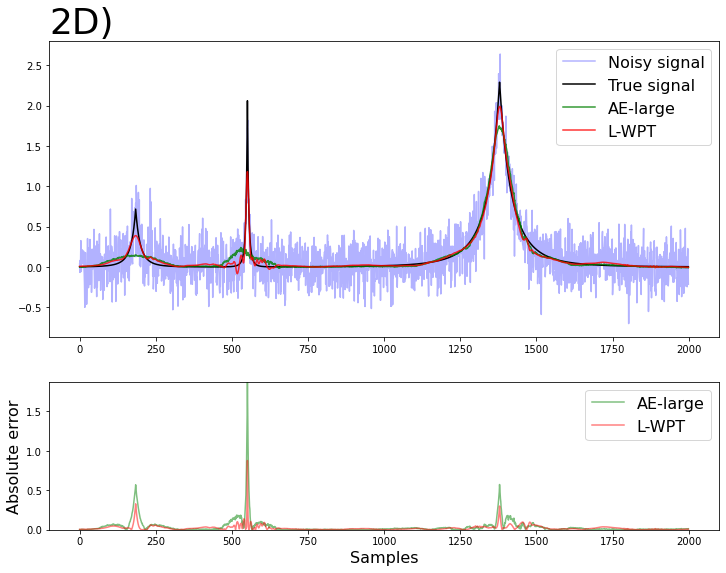}
\includegraphics[width=0.24\textwidth]{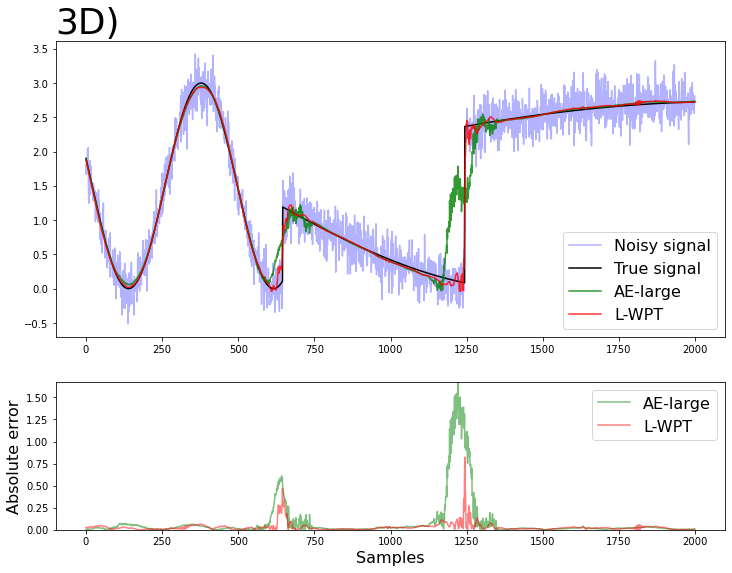}
\includegraphics[width=0.24\textwidth]{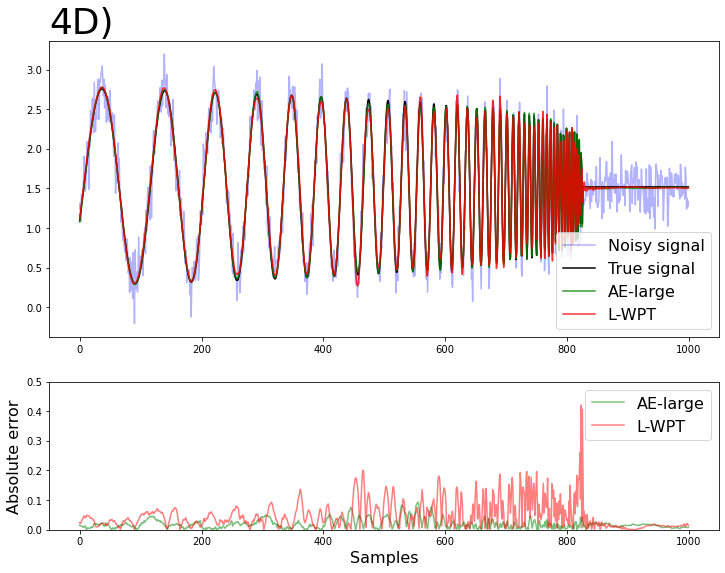}
\caption{Table of figures showing denoising examples for the AE-large and L-WPT methods. The numbers one to four and the letter A to D represent respectively the Block, Bumps, HeaviSine and Doppler classes, the number is the represented class where the letter is the class used to train the model.}\label{fig-3C2}
\end{figure*}

The Figure~\ref{fig-3C2} can be seen as a table of figures, where the columns provide a realisation of each class, and the rows provide the output of the L-WPT and AE big when they are trained with one of the four classes. For each figure, the absolute error between the estimated and the pure signal through time is provided.
AE-large performs the best when applied to the training class (the figures on the diagonal). However it performs poorly when it is trained using another class. On the contrary, the L-WPT is more consistent regarding if it is applied on the training class or not.

\begin{table*}
\center
\begin{tabular}{c|ccc|ccc|ccc|ccc||ccc}
\hline
\multirow{2}{*}{Method} &    \multicolumn{3}{c|}{Block} &    \multicolumn{3}{c|}{Bump}    &    \multicolumn{3}{c|}{HeaviSine} &    \multicolumn{3}{c||}{Doppler} &    \multicolumn{3}{c}{\textbf{Mean results}}   \\ & $S_p$ &  $S_r$ & $\bar{S} $ &  $S_p$ &  $S_r$ & $\bar{S} $ &  $S_p$ &  $S_r$ & $\bar{S} $ &  $S_p$ &  $S_r$ & $\bar{S} $&  $S_p$ &  $S_r$ & $\bar{S} $  \\
\hline 
Baseline-HT& 354& \textbf{246}& 273& 194& \textbf{300}& \textbf{273}& 225& 291& 274& 317& 260& 275& 273& 274& 274\\
AE-large& \textbf{24}& 577& 439& \textbf{20}& 1859& 1399& \textbf{33}& 563& 431& \textbf{36}& 1575& 1190& \textbf{28}& 1143& 865\\
AE-small& 127& 332& 281& 56& 706& 544& 129& 383& 319& 233& 226& 228& 136& 412& 343\\
CNN-large& 75& 355& 285& 39& 965& 734& 75& 355& 285& 94& 305& 252& 71& 495& 389\\
CNN-small& 189& 329& 294& 79& 624& 488& 195& 342& 305& 270& 259& 262& 183& 389& 337\\
Unet-large & 27& 423& 324& 21& 1508& 1136& 35& 309& 240& 36& 810& 617& 30& 763& 579\\
Unet-small& 136& 339& 288& 61& 646& 499& 133& 352& 297& 201& 230& 223& 133& 392& 327\\
L-WPT& 83& 315& \textbf{257}& 43& 358& 279& 56& \textbf{195}& \textbf{161}& 98& \textbf{143}& \textbf{132}& 70& \textbf{253}& \textbf{207}\\
\hline
\end{tabular} 
\caption{specialisation score ($S_p$), robustness score ($S_r$) and ($\bar{S} $) mean score over the 4 function types when methods are trained using only the classes of the corresponding column.  }\label{TableC}
\end{table*}



\subsection{Impact of the noise level}\label{p5C}
In practical applications, noise levels can change over time. This can for example occur under new operating conditions or if the training was done combining signals and background with specified SNR \cite{nossier2020experimental}. In our setup, the training noise level is fixed to $\sigma=0.2$. We then quantify the performance of each method for different values of the noise level.

The $S_p$, $S_r$ and $\bar{S}$ scores are computed again when the noise level takes the values $\{0.1, 0.4, 0.6, 0.8, 1\}$. Since the thresholds learned for the L-WPT are fixed to perform well for $\sigma=0.2$, there is no particular reason that it will continue to do a relevant denoising for other noise levels. To adjust the weights, we perform the L-WPT-($\delta$) transformation introduced in Section~\ref{p2F} with $\delta=\frac{\sigma}{ \sigma_{\rm train}}=5\sigma$. Here, we assume we have a good estimation of the noise level for the new operating condition. In order to highlight the denoising performance of our L-WPT-($\delta$) method, we compute again the best threshold of the Baseline-HT method for each noise level.  

\begin{figure}
\centering
\includegraphics[width=0.23\textwidth]{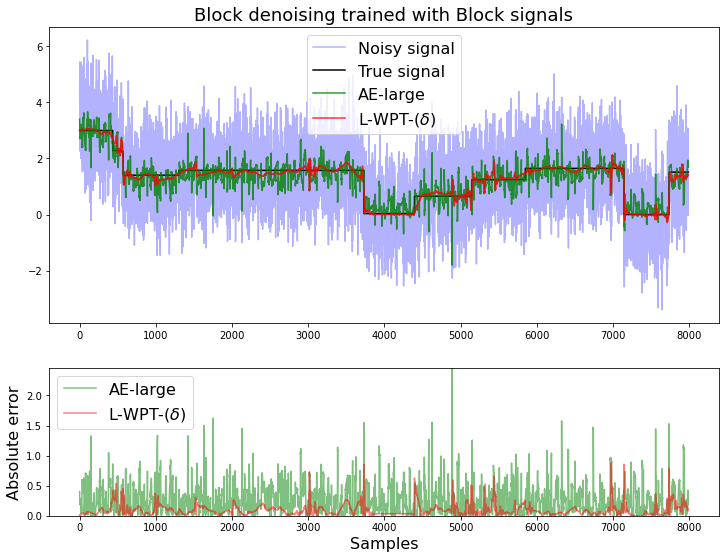}
\includegraphics[width=0.23\textwidth]{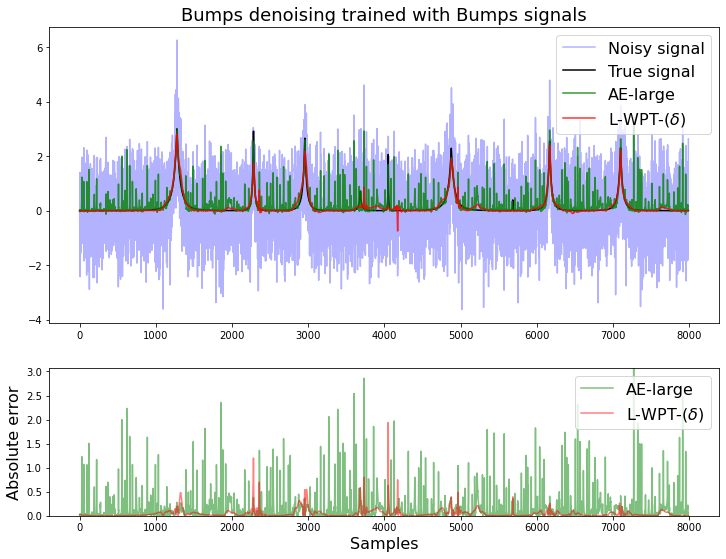}

\includegraphics[width=0.23\textwidth]{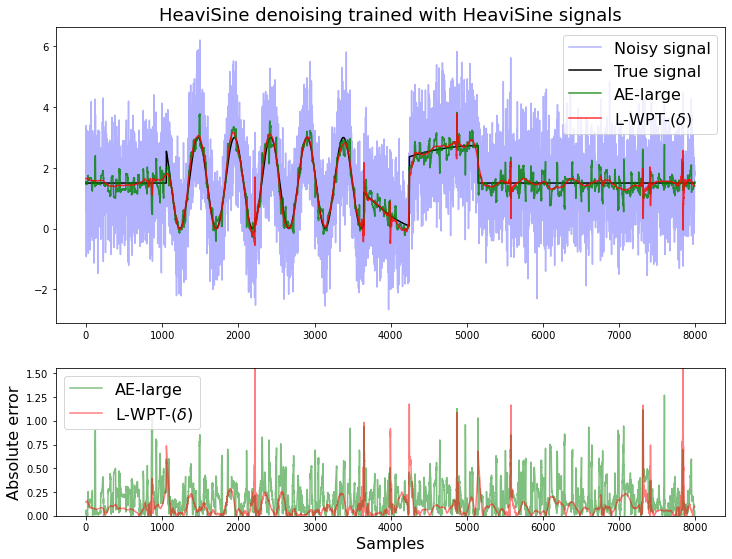}
\includegraphics[width=0.23\textwidth]{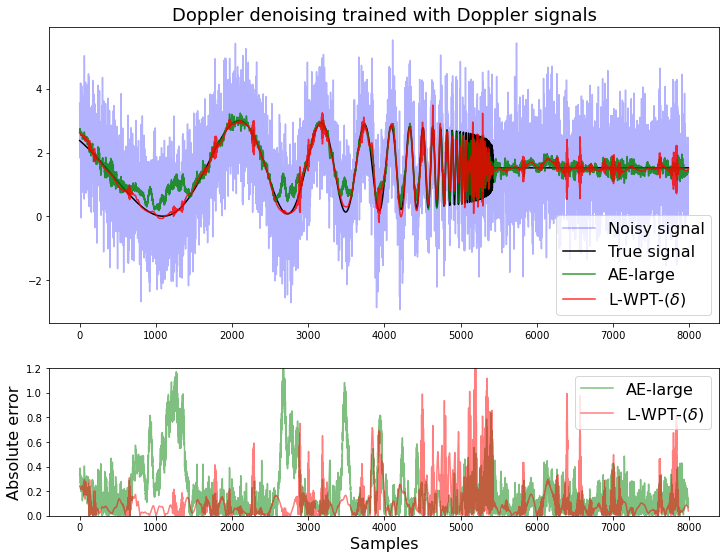}
\caption{(Top) Denoising with L-WPT (weight mult) and AE-large when they are trained using the corresponding class function with a noise level of $\sigma=0.2$ and tested on a noise level of $\sigma=1$. (Bottom) Abslolute error between the denoised signal and the ground truth}\label{fig-3D1}
\end{figure}

Figure~\ref{fig-3D1} shows the specialisation and the robustness scores for each method, for the different noise levels and for each training class. We display the decimal logarithmic value of the scores in order to ease the reading of the graphs. The L-WPT performs poorly. However, the modified version with the weight adjustment outperforms all other methods. In the case of $\sigma=1$, the L-WPT-($5\sigma$) method can provide an up to 10-times better denoising capability compared to the deep NN models. The L-WPT-($\delta$) also outperforms the Baseline-HT method where the threshold was optimised to the new level of noise. It demonstrates that the filters learned by L-WPT are robust to higher levels of noise and that only the biases need to be adjusted. 

The Figure~\ref{fig-3D1} demonstrates the denoising performance of L-WPT-($5\sigma$) compared to the AE-large method for the test signal with $\sigma=1$ (both methods are trained using the class on which they are also tested: specialization regime).

\begin{figure}
\centering
\includegraphics[width=0.45\textwidth]{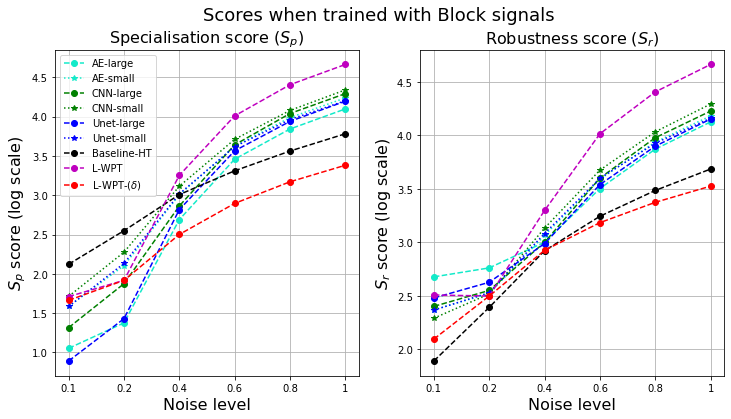}

\includegraphics[width=0.45\textwidth]{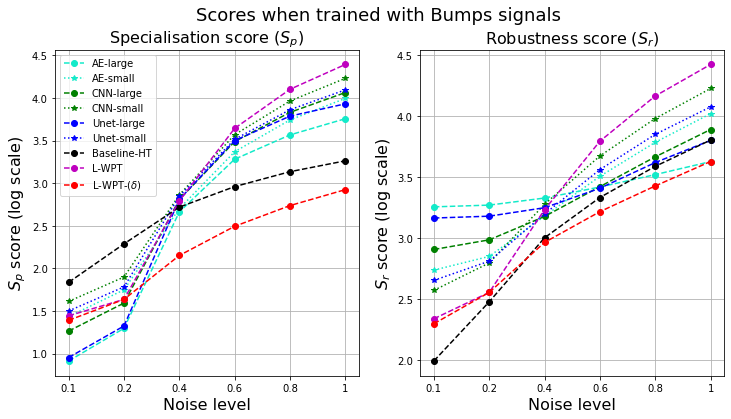}

\includegraphics[width=0.45\textwidth]{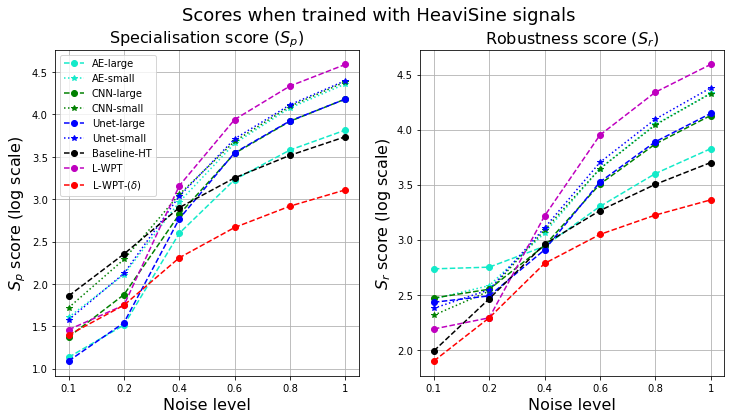}

\includegraphics[width=0.45\textwidth]{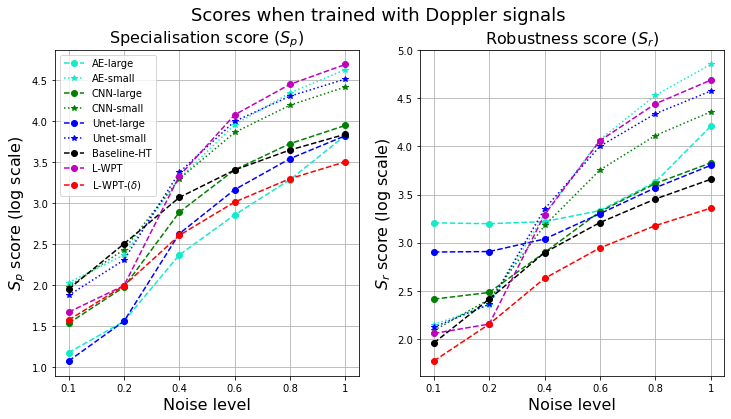}
\caption{Logarithm of the $S_p$ and $S_r$ scores through different noise level for each method when trained using a noise level of $\sigma = 0.2$.}\label{fig-3D2}
\end{figure}

\section{Application on audio background denoising}\label{Section4}

We evaluate the proposed L-WPT on denoising real signals. For this case study, we also compare to the same methods presented in subsection~\ref{pComp}. The alternative methods are the Baseline-HT, AE-large, AE-small, CNN-large, CNN-small, UNet-large and UNet-small. For the deep NNs models, we use the same architecture as reported in Table~\ref{Table3_1}. The justification of this choice is provided in Appendix~\ref{AppendixB}. 

After presenting the dataset in subsection~\ref{ss1} the denoising performance of each method is reported in subsection~\ref{ss3} with respect to the robustness and specification score. For this case, the SNR is known and an application on real conditions with unknown SNR is provided in subsection~\ref{ss4}. We provide in this subsection a method to estimate the $\delta$ value of the $\delta$ transform. Finally an analysis of the trained L-WPT for the airport background removal is provided in subsection~\ref{ss2}.

\subsection{Dataset}\label{ss1}

We consider the dataset from the task 1 and 2 of the DCASE 2018 challenge \cite{mesaros2018multi}, \cite{fonseca2018general}. The first dataset provides acoustic scenes that are used as noisy backgrounds. We only consider the scenes of the airports of Barcelona, Helsinki, London, Paris and Stockholm. The second task provides a variety of 41 different foreground sound events like "Cello", "Bus" or "Bark". We randomly eliminated the "Trumpet" classes in order to keep only 40 different classes and ease the division of the dataset into folds. We consider only signals where the classe label have been checked manually.

The background training signals are based on 102 10-seconds recordings of the Barcelona airport scene only, the test background uses 26 different recordings of the Barcelona airport or the recordings of the other airport scenes. Moreover, the foreground training signals are based on 3610 recordings, different from the 1600 recordings used for the test classes. 

For the audio signal generation of the foreground and background signals, we apply a similar strategy as in \cite{nossier2020experimental}. The recordings are downsampled to 8 kHz, cropped randomly in order to have signals with $T=2^{13}$ samples, and normalized. For the foreground sound, we apply padding and make sure that the random cropping does not select a null signal. We mix the foreground and background sound by adding them, the signal to noise ratio in this case is 0 DB. For the training, new signals are continuously generated from the training recordings. 1600 test signals are generated from the test recordings
 
The methods are not trained on all the classes directly, the data is cut into 8 folds of 5 classes each. This aims to mimic real world applications: A limited amount of classes is collected and used for training the model. The trained model is then applied in a more general environment where we aim to eliminate the background also for classes of signals different from the training data.

\subsection{Robust background removal}\label{ss3}

We consider the L-WPT with eight layers. It corresponds to the number of layers minimising the entropy for the WPT when applied to the pure signals of the first fold, the entropy minimisation is a standard method used to select the best number of layers \cite{coifman1992entropy}.

The $S_p$, $S_r$ and $\bar{S}$ scores introduced in Section~\ref{p4C} are also applied in this case study, where the training classes $C_{\rm train}$ are the 5 classes of the current fold. We also consider the mean square error between the estimated and pure signals obtained for each class separately. 

\begin{figure}
\centering
\includegraphics[width=0.49\textwidth]{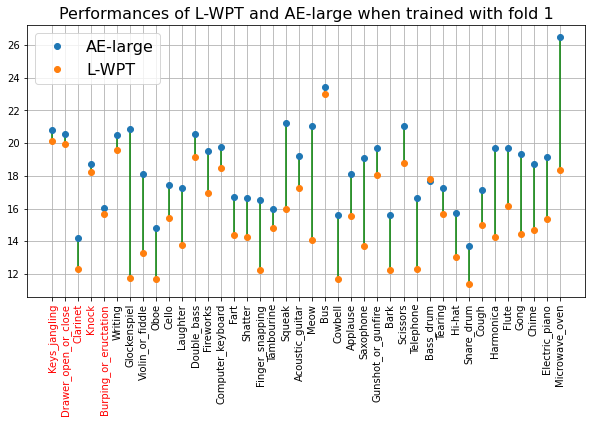}
\includegraphics[width=0.49\textwidth]{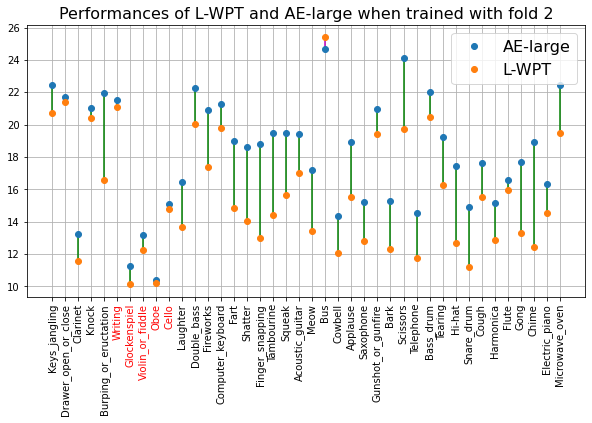}
\caption{Performances of the AE-large and L-WPT for each classes when they are trained using three different folds. The classes belonging to $C_{\rm train}$ are in red, the difference between the score obtained via L-WPT and AE-large is highlighted in green when L-WPT works best, in purple when AE-large works best.}\label{fig-BackBar}
\end{figure}

Figure~\ref{fig-BackBar} shows the normalised mean square error obtained for each class when the L-WPT and the AE-large were trained with the fold 1 and 2. The classes belonging to $C_{\rm train}$ are displayed in red, the difference between the scores obtained with L-WPT and AE-large is highlighted in green when L-WPT performs best and in purple when AE-large performs best.
For this experiment, the L-WPT almost always outperforms the AE-large.
The gap between the L-WPT and the AE-large MSE is reduced for $C_{\rm train}$. For example in the fold 2, the MSE is reduced by 2.6 on average for all classes, whereas it is reduced only by 0.6 if we only consider the training class. It shows how our method is able to generalize well to structured signal from classes different from the training dataset. 

\begin{figure}
\centering
\includegraphics[width=0.23\textwidth]{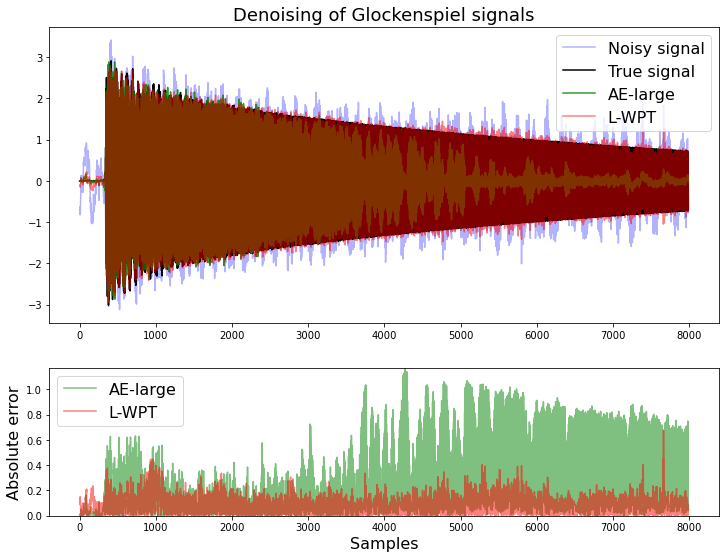}
\includegraphics[width=0.23\textwidth]{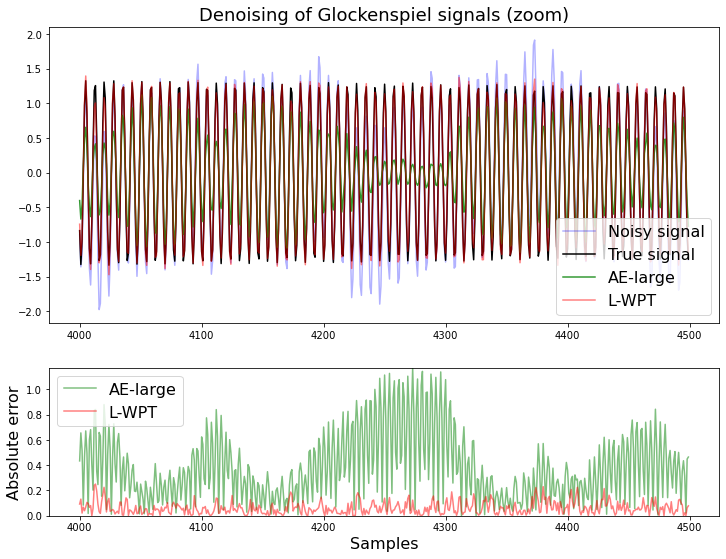}

\includegraphics[width=0.23\textwidth]{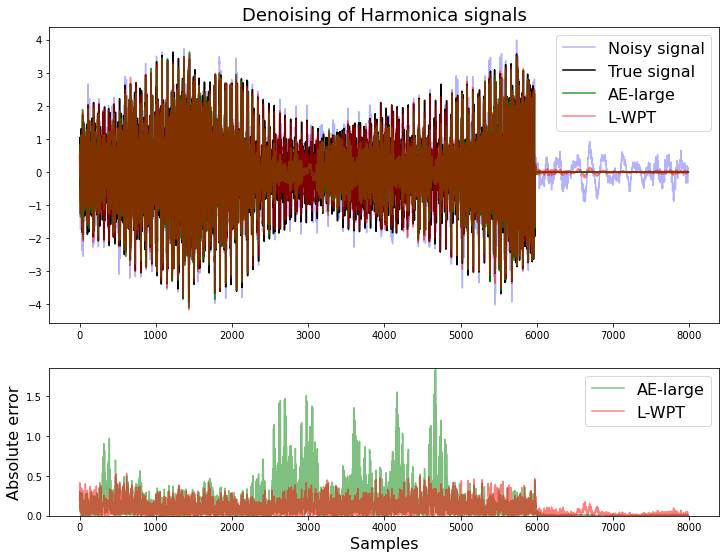}
\includegraphics[width=0.23\textwidth]{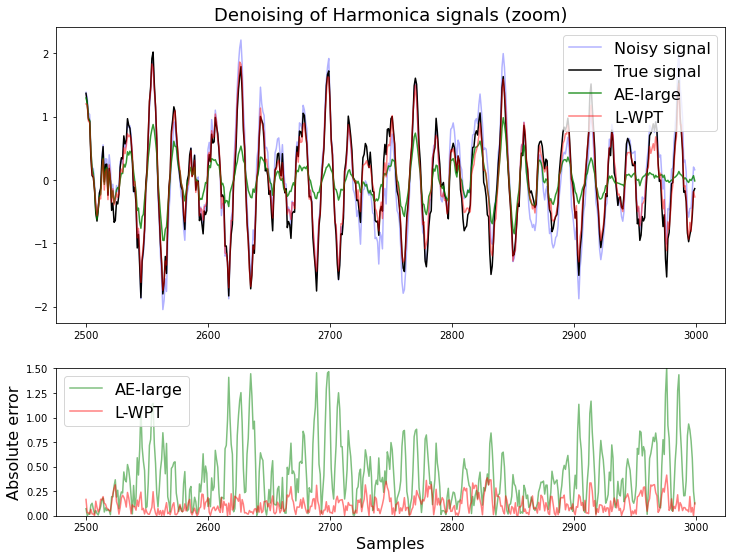}

\includegraphics[width=0.23\textwidth]{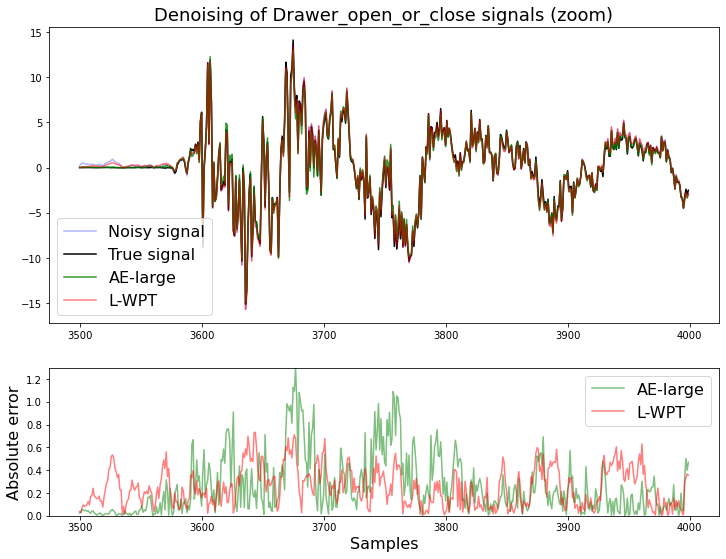}
\includegraphics[width=0.23\textwidth]{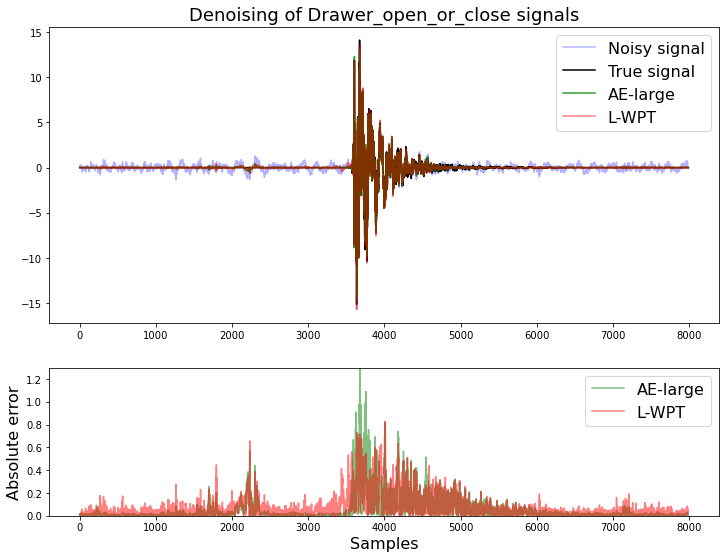}
\caption{(Top) Denoising with L-WPT (weight mult) and AE-large when they are trained using the fold one. (Bottom) Abslolute error between the denoised signal and the ground truth}\label{fig-BackBar2}
\end{figure}

Figure~\ref{fig-BackBar2} shows an example of denoising with L-WPT and AE-large when they where trained with the fold 1. The two first cases, "Glockenspiel" and "Harmonica" signals, are cases where the L-WPT performs particularly well compared to the AE-large. The last case "Drawer open and close" is a case where both methods perform similarly. In this last case, the AE-large performs better in eliminating noise alone, however, the L-WPT is slightly more accurate in reconstructing the sound of interest (as indicated by the absolute error).  

In Table~\ref{TableX}, the mean $S_p$, $S_r$ and $\bar{S}$ scores over the 8 folds and for each method are provided. The score is obtained for the Bacelona airport scene background sounds and the other airport sounds. We recall that the training data use only background sounds from the Barcelona airport. Because the background noise contains specific frequency contents, the baseline-HT method is no longer adapted since it denoises each frequency bands in the same way. It, therefore, gives poor results. Overall, the L-WPT outperforms all other methods. For the airport case, the gap between the specialisation score and the robustness score is 0.5 which is low. However, the AE-large has a gap between the specialisation score and the robustness
score of 2.4 which is comparable large. This demonstrates again that L-WPT has the learning capabilities of deep NNs while keeping the universal properties of signal processing.
Considering the background of other airports, the trend is similar, with L-WPT outperforming the other methods in terms of specialization and robustness scores. It also shows that there are not many differences between the background sounds of different airports.

\begin{table*}
\center
\begin{tabular}{c|ccc||ccc|ccc|ccc|ccc}
\hline
Method &    \multicolumn{3}{c||}{Barcelona} &    \multicolumn{3}{c|}{Helsinky}    &    \multicolumn{3}{c|}{London} &    \multicolumn{3}{c|}{Paris} &    \multicolumn{3}{c}{Stokholm}   \\ & $S_p$ &  $S_r$ & $\bar{S} $ &  $S_p$ &  $S_r$ & $\bar{S} $ &  $S_p$ &  $S_r$ & $\bar{S} $ &  $S_p$ &  $S_r$ & $\bar{S} $&  $S_p$ &  $S_r$ & $\bar{S} $  \\
\hline 
Baseline-HT& 28.2& 28.4& 28.4& 28.3& 28.4& 28.4& 28.1& 28.2& 28.2& 28.4& 28.2& 28.2& 28.1& 28.1& 28.1\\
AE-large& 14.9& 17.3& 17.0& 14.8& 17.3& 17.0& 15.8& 18.0& 17.7& 15.2& 17.6& 17.3& 15.8& 18.1& 17.8\\
AE-small& 21.3& 22.6& 22.5& 20.7& 22.1& 21.9& 21.6& 22.9& 22.7& 21.2& 22.5& 22.3& 21.4& 22.7& 22.5\\
CNN-large& 16.7& 18.3& 18.1& 16.4& 18.0& 17.8& 17.2& 18.7& 18.5& 16.8& 18.4& 18.2& 17.0& 18.5& 18.4\\
CNN-small& 18.3& 18.9& 18.8& 17.6& 18.3& 18.2& 18.4& 19.0& 18.9& 18.0& 18.7& 18.6& 18.2& 18.9& 18.8\\
Unet-large & 16.2& 19.0& 18.7& 16.1& 19.0& 18.6& 17.2& 19.8& 19.5& 16.6& 19.4& 19.0& 17.1& 19.8& 19.4\\
Unet-small& 17.6& 18.3& 18.2& 16.9& 17.6& 17.5& 17.9& 18.5& 18.4& 17.4& 18.1& 18.0& 17.6& 18.3& 18.2\\
L-WPT& \textbf{14.3}& \textbf{14.8}& \textbf{14.8}& \textbf{13.7}& \textbf{14.3}& \textbf{14.2}& \textbf{15.1}& \textbf{15.6}& \textbf{15.6}& \textbf{14.3}& \textbf{14.8}& \textbf{14.8}& \textbf{15.1}& \textbf{15.6}& \textbf{15.6}\\
\hline
\end{tabular} 
\caption{specialisation score ($S_p$), robustness score ($S_r$) and mean score ($\bar{S} $) over the 8 folds when each method are trained with the Barcelona noise only.  }\label{TableX}
\end{table*}

\subsection{Real conditions application with unknown SNR}\label{ss4}
Each method was trained based on corrupted signals with a fixed SNR. However, the denoising performances can decrease if applied in real applications to sounds with different SNR. Indeed, depending on the location of the sensor in the airport or the recording time, the SNR can differ significantly. Thus, we now consider different SNR for the test signal.
For this, we use the non-normalised airport background noises. Since their raw value ranges are too low compared to the signals of interest, we multiplied them by 200. These values are chosen so that the majority of signals has a higher SNR than for the training case. This corresponds to the situation where the denoising can be impaired.

We impose that the first 2000 samples of the test signals contain only the background noise. Thereby, it is possible to evaluate the $\delta$ value for the L-WPT-$\delta$ transformation. The $\delta$ value is defined as the fraction between the norm of the 2000 first samples over the average norm when we select randomly 2000 samples from the background of the training dataset. The idea is to see if the background energy of the current recording is higher or lower than from the training dataset.

\begin{figure}[h]
\centering
\includegraphics[width=0.49\textwidth]{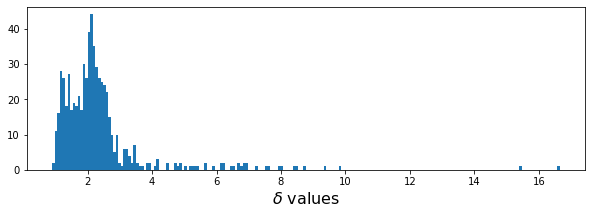}
\caption{Frequency of $\delta$ values modifying the biase of the L-WPT-($\delta$) method over the test dataset}\label{fig-BackBar3}
\end{figure}

Figure~\ref{fig-BackBar3} shows the histograms of the obtained $\delta$ values over the 1600 recordings of the test dataset. On average, the value of $\delta$ is larger than one. This means that the SNR for the test dataset is negative. Table~\ref{TableY} shows the $S_p$, $S_r$ and $\bar{S}$ results for each of the applied deep NN architectures. The extended L-WPT-$\delta$  outperforms other methods with respect to each of the scores. It shows again how the L-WPT learn kernels that are robust to different noise level. Thus, the L-WPT can be easily adapted to different operating conditions by modifying the biases only. 

\begin{table}
\center
\begin{tabular}{c|ccc}
\hline
Method &    \multicolumn{3}{c}{Barcelona}   \\ & $S_p$ &  $S_r$ & $\bar{S} $   \\
\hline 
AE-large& 19.3& 21.5& 21.2\\
AE-small& 26.2& 27.4& 27.3\\
CNN-large& 21.6& 23.1& 22.9\\
CNN-small& 23.3& 23.9& 23.8\\
Unet-large & 19.8& 22.6& 22.2\\
Unet-small& 22.8& 23.4& 23.3\\
L-WPT-($\delta$)& \textbf{18.5}& \textbf{18.9}& \textbf{18.9}\\
\hline
\end{tabular} 
\caption{specialisation score ($S_p$), robustness score ($S_r$) and mean score ($\bar{S} $) over the 8 folds when each method are trained with the Barcelona noise at 0db, the test dataset has varying noise SNR.  }\label{TableY}
\end{table}

\subsection{Details of the L-WPT training}\label{ss2}

\begin{figure*}
\centering
\includegraphics[width=0.23\textwidth]{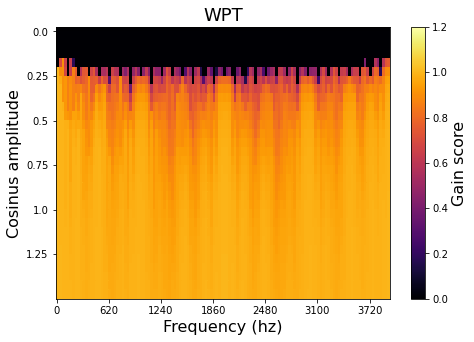}
\includegraphics[width=0.23\textwidth]{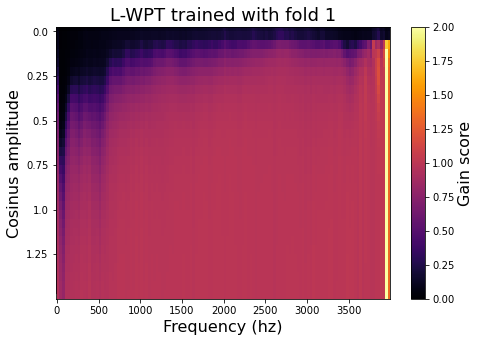}
\includegraphics[width=0.23\textwidth]{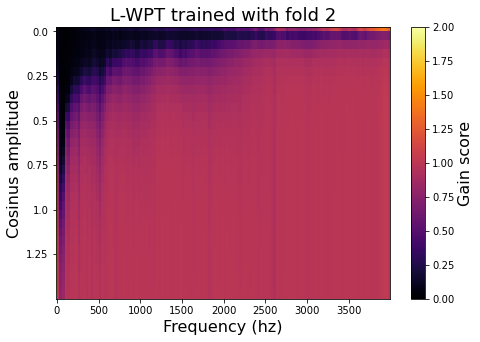}
\includegraphics[width=0.23\textwidth]{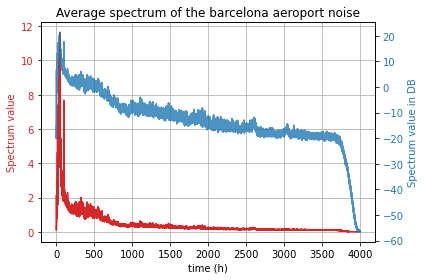}
\caption{(left) Gain score for the WPT with HT denoising considering a thresholding value $\lambda=1$ (Middle) Gain score from pure cosines signal with increasing frequency and amplitude for the L-WPT trained respectively on the first two folds. (right) Mean spectrogram (modulus and decibels) of the barcelona airport noise signals}\label{fig-4B1}
\end{figure*}

In this section we propose to compare the filtering provided by L-WPT and baseline-HT in the context of airport background noise removal. The impact of the baseline-HT denoising strategy on cosines of different frequencies and amplitudes is presented in the left Figure~\ref{fig-4B1}. It shows the gain score, which is the ratio of the norm of the input cosine and the norm of the output signal, in function of the amplitude and the frequency of the input.
The frequency range goes from 0 Hz to the maximum frequency (2**12 Hz), and the amplitude range goes from 0 to 1.5.
When cosines have a too low amplitude, they are interpreted as noise and are not reconstructed, which corresponds to a gain score of 0. On the contrary, cosines with a high amplitude are perfectly reconstructed and have a score of 1. Some imperfections are present since the applied filters are not ideal. 

The two middle Figure~\ref{fig-4B1} visualizes the gain scores for the two first folds. We can see that even if each L-WPT was trained using the signals from different classes, the gain scores images appear to be relatively similar. In comparison the the Baseline-HT method, the denoising is adapted to the signal frequency.
The right Figure~\ref{fig-4B1} shows the average spectrum (in absolute values and in Decibels (DB)) of the training background noise. We can see that it contains mostly low frequency contents from 0 to 800 Hz. It is interesting to remark that the gain score for cosines with a low frequency content from 0 to 800 Hz stays null for higher amplitudes than cosines with a higher frequency. It shows how the L-WPT learned to suppress the background contents. For the highest frequencies ($>$3500), the background content is almost null and it turns out that the gain scores are more heterogeneous from one fold to another. It implies the L-WPT were more able to specialise to the training fold for non-corrupted frequency bands. 

\section{Conclusion}
In this paper, we propose to combine two of the main signal denoising tools: Wavelet shrinkage using wavelet packet transform and supervised denoising by a convolutional autoencoder. 
Our proposed learnable WPT is interpretable, relying on the signal processing properties of WPT while being able to learn the specifics of the training dataset. 
Moreover, it is able to generalize to different classes of the training dataset. 
It has an intuitive parameter initialization that allows it to initialize like a wavelet packet transform.  
Moreover, we propose a powerful post-learning modification of the weights, called the $\delta$-modification. This modification is only possible because the meaning of each parameter in this architecture is known. 
Thus, it is possible to adapt denoising to different noise levels resulting from different operating conditions. 

The L-WPT is compared to deep supervised models and the WPT denoising in two experiments.
It was first applied on case functions often used in the denoising literature. 
The L-WPT was able to learn specific denoising for the signals in the training class. 
Furthermore, we demonstrate that it retains the robustness of a universal signal processing procedure by testing it on noisy signals outside the training class. 
We also show that our method was also robust to different types and levels of noise thanks to the $\delta$-modification. 
Finally, the L-WPT method was applied to a background suppression task and performed better than the other methods. 
We provide a recommendation for using the $\delta$-modification in a real application that has been shown to be effective for background denoising in a variable SNR.

This work opens several doors for future directions. 
First, further research on the $\delta$-modification or related modifications should be conducted. 
For example, learning the L-WPT on different noise levels would show if the kernels remain similar.This would tell us how optimal the $delta$-modification is. 
Another future direction would be to use the time-frequency representation of L-WPT as a feature for a supervised task instead of the WPT features.
On the application side, it would be interesting to apply our approach in the context of speech enhancement with a fixed background. 
Finally, the generalization of our approach to  multi-dimensional signals would lead to its application to image denoising.

\section*{Aknowledgments}
This study was financed by the Swiss Innovation Agency (lnnosuisse) under grant number: 47231.1 IP-ENG.

\bibliographystyle{elsarticle-num}

\bibliography{bibsample}

\begin{thebibliography}{10}
\expandafter\ifx\csname url\endcsname\relax
  \def\url#1{\texttt{#1}}\fi
\expandafter\ifx\csname urlprefix\endcsname\relax\def\urlprefix{URL }\fi
\expandafter\ifx\csname href\endcsname\relax
  \def\href#1#2{#2} \def\path#1{#1}\fi

\bibitem{li2009image}
H.~Li, F.~Liu, Image denoising via sparse and redundant representations over
  learned dictionaries in wavelet domain, in: 2009 Fifth International
  Conference on Image and Graphics, IEEE, 2009, pp. 754--758.

\bibitem{li2018data}
Y.~Li, C.~Xu, L.~Yi, R.~Fang, A data-driven approach for denoising gnss
  position time series, Journal of Geodesy 92~(8) (2018) 905--922.

\bibitem{jha2010denoising}
S.~K. Jha, R.~Yadava, Denoising by singular value decomposition and its
  application to electronic nose data processing, IEEE Sensors Journal 11~(1)
  (2010) 35--44.

\bibitem{rajwade2012image}
A.~Rajwade, A.~Rangarajan, A.~Banerjee, Image denoising using the higher order
  singular value decomposition, IEEE Transactions on Pattern Analysis and
  Machine Intelligence 35~(4) (2012) 849--862.

\bibitem{frusque2021canonical}
G.~Frusque, M.~Gabriel, F.~Olga, Canonical polyadic decomposition and deep
  learning for machine fault detection, in: PHM Society European Conference,
  Vol.~6, 2021, pp. 9--9.

\bibitem{bayer2019iterative}
F.~M. Bayer, A.~J. Kozakevicius, R.~J. Cintra, An iterative wavelet threshold
  for signal denoising, Signal Processing 162 (2019) 10--20.

\bibitem{gharesi2020neuro}
N.~Gharesi, M.~M. Arefi, R.~Razavi-Far, J.~Zarei, S.~Yin, A neuro-wavelet based
  approach for diagnosing bearing defects, Advanced Engineering Informatics 46
  (2020) 101172.

\bibitem{lv2021predictive}
Y.~Lv, Q.~Zhou, Y.~Li, W.~Li, A predictive maintenance system for
  multi-granularity faults based on adabelief-bp neural network and fuzzy
  decision making, Advanced Engineering Informatics 49 (2021) 101318.

\bibitem{zhao2022highly}
M.~Zhao, X.~Fu, Y.~Zhang, L.~Meng, B.~Tang, Highly imbalanced fault diagnosis
  of mechanical systems based on wavelet packet distortion and convolutional
  neural networks, Advanced Engineering Informatics 51 (2022) 101535.

\bibitem{donoho1994ideal}
D.~L. Donoho, J.~M. Johnstone, Ideal spatial adaptation by wavelet shrinkage,
  biometrika 81~(3) (1994) 425--455.

\bibitem{zhou2020partial}
S.~Zhou, J.~Tang, C.~Pan, Y.~Luo, K.~Yan, Partial discharge signal denoising
  based on wavelet pair and block thresholding, IEEE Access 8 (2020)
  119688--119696.

\bibitem{kumar2021stationary}
A.~Kumar, H.~Tomar, V.~K. Mehla, R.~Komaragiri, M.~Kumar, Stationary wavelet
  transform based ecg signal denoising method, ISA transactions 114 (2021)
  251--262.

\bibitem{alyasseri2019eeg}
Z.~A.~A. Alyasseri, A.~T. Khader, M.~A. Al-Betar, A.~K. Abasi, S.~N. Makhadmeh,
  Eeg signals denoising using optimal wavelet transform hybridized with
  efficient metaheuristic methods, IEEE Access 8 (2019) 10584--10605.

\bibitem{chang2000adaptive}
S.~G. Chang, B.~Yu, M.~Vetterli, Adaptive wavelet thresholding for image
  denoising and compression, IEEE transactions on image processing 9~(9) (2000)
  1532--1546.

\bibitem{zhang2017beyond}
K.~Zhang, W.~Zuo, Y.~Chen, D.~Meng, L.~Zhang, Beyond a gaussian denoiser:
  Residual learning of deep cnn for image denoising, IEEE transactions on image
  processing 26~(7) (2017) 3142--3155.

\bibitem{nossier2020comparative}
S.~A. Nossier, J.~Wall, M.~Moniri, C.~Glackin, N.~Cannings, A comparative study
  of time and frequency domain approaches to deep learning based speech
  enhancement, in: 2020 International Joint Conference on Neural Networks
  (IJCNN), IEEE, 2020, pp. 1--8.

\bibitem{xie2021bioacoustic}
J.~Xie, J.~G. Colonna, J.~Zhang, Bioacoustic signal denoising: a review,
  Artificial Intelligence Review 54~(5) (2021) 3575--3597.

\bibitem{arsene2019deep}
C.~T. Arsene, R.~Hankins, H.~Yin, Deep learning models for denoising ecg
  signals, in: 2019 27th European Signal Processing Conference (EUSIPCO), IEEE,
  2019, pp. 1--5.

\bibitem{kounovsky2017single}
T.~Kounovsky, J.~Malek, Single channel speech enhancement using convolutional
  neural network, in: 2017 IEEE International Workshop of Electronics, Control,
  Measurement, Signals and their Application to Mechatronics (ECMSM), IEEE,
  2017, pp. 1--5.

\bibitem{liu2019fault}
X.~Liu, Q.~Zhou, J.~Zhao, H.~Shen, X.~Xiong, Fault diagnosis of rotating
  machinery under noisy environment conditions based on a 1-d convolutional
  autoencoder and 1-d convolutional neural network, Sensors 19~(4) (2019) 972.

\bibitem{nossier2020experimental}
S.~A. Nossier, J.~Wall, M.~Moniri, C.~Glackin, N.~Cannings, An experimental
  analysis of deep learning architectures for supervised speech enhancement,
  Electronics 10~(1) (2020) 17.

\bibitem{lyu2022novel}
P.~Lyu, K.~Zhang, W.~Yu, B.~Wang, C.~Liu, A novel rsg-based intelligent bearing
  fault diagnosis method for motors in high-noise industrial environment,
  Advanced Engineering Informatics 52 (2022) 101564.

\bibitem{ravanelli2018speaker}
M.~Ravanelli, Y.~Bengio, Speaker recognition from raw waveform with sincnet,
  in: 2018 IEEE Spoken Language Technology Workshop (SLT), IEEE, 2018, pp.
  1021--1028.

\bibitem{frusque2022learnable}
F.~Ga{\"e}tan, F.~Olga, Learnable wavelet packet transform for data-adapted
  spectrograms, in: ICASSP 2022-2022 IEEE International Conference on
  Acoustics, Speech and Signal Processing (ICASSP), IEEE, 2022, pp. 3119--3123.

\bibitem{michau2021fully}
G.~Michau, O.~Fink, Fully learnable deep wavelet transform for unsupervised
  monitoring of high-frequency time series, arXiv preprint arXiv:2105.00899.

\bibitem{oktar2016speech}
M.~A. Oktar, M.~Nibouche, Y.~Baltaci, Speech denoising using discrete wavelet
  packet decomposition technique, in: 2016 24th Signal Processing and
  Communication Application Conference (SIU), IEEE, 2016, pp. 817--820.

\bibitem{kumar2015comparative}
G.~Kumar, S.~Kumar, N.~Kumar, Comparative study of wavelet and wavelet packet
  transform for denoising telephonic speech signal, International Journal of
  Computer Applications 110~(15).

\bibitem{schimmack2016noise}
M.~Schimmack, P.~Mercorelli, Noise detection for biosignals using orthogonal
  wavelet packet tree denoising algorithm, International Journal of Electronics
  and Telecommunications 62~(1) (2016) 15--21.

\bibitem{schimmack2018wavelet}
M.~Schimmack, P.~Mercorelli, A wavelet packet tree denoising algorithm for
  images of atomic-force microscopy, Asian Journal of Control 20~(4) (2018)
  1367--1378.

\bibitem{beale2020adaptive}
C.~Beale, C.~Niezrecki, M.~Inalpolat, An adaptive wavelet packet denoising
  algorithm for enhanced active acoustic damage detection from wind turbine
  blades, Mechanical Systems and Signal Processing 142 (2020) 106754.

\bibitem{yue2019bayesian}
G.-d. Yue, X.-s. Cui, Y.-y. Zou, X.-t. Bai, Y.-H. Wu, H.-t. Shi, A bayesian
  wavelet packet denoising criterion for mechanical signal with non-gaussian
  characteristic, Measurement 138 (2019) 702--712.

\bibitem{fu2017raw}
S.-W. Fu, Y.~Tsao, X.~Lu, H.~Kawai, Raw waveform-based speech enhancement by
  fully convolutional networks, in: 2017 Asia-Pacific Signal and Information
  Processing Association Annual Summit and Conference (APSIPA ASC), IEEE, 2017,
  pp. 006--012.

\bibitem{kong2019sound}
Q.~Kong, Y.~Xu, I.~Sobieraj, W.~Wang, M.~D. Plumbley, Sound event detection and
  time--frequency segmentation from weakly labelled data, IEEE/ACM Transactions
  on Audio, Speech, and Language Processing 27~(4) (2019) 777--787.

\bibitem{kachuee2018dynamic}
M.~Kachuee, S.~Darabi, B.~Moatamed, M.~Sarrafzadeh, Dynamic feature acquisition
  using denoising autoencoders, IEEE transactions on neural networks and
  learning systems 30~(8) (2018) 2252--2262.

\bibitem{grozdic2017whispered}
D.~T. Grozdi{\'c}, S.~T. Jovi{\v{c}}i{\'c}, M.~Suboti{\'c}, Whispered speech
  recognition using deep denoising autoencoder, Engineering Applications of
  Artificial Intelligence 59 (2017) 15--22.

\bibitem{fink2020potential}
O.~Fink, Q.~Wang, M.~Svensen, P.~Dersin, W.-J. Lee, M.~Ducoffe, Potential,
  challenges and future directions for deep learning in prognostics and health
  management applications, Engineering Applications of Artificial Intelligence
  92 (2020) 103678.

\bibitem{lv2022vibration}
Y.~Lv, W.~Zhao, Z.~Zhao, W.~Li, K.~K. Ng, Vibration signal-based early fault
  prognosis: Status quo and applications, Advanced Engineering Informatics 52
  (2022) 101609.

\bibitem{berghout2020aircraft}
T.~Berghout, L.-H. Mouss, O.~Kadri, L.~Sa{\"\i}di, M.~Benbouzid, Aircraft
  engines remaining useful life prediction with an adaptive denoising online
  sequential extreme learning machine, Engineering Applications of Artificial
  Intelligence 96 (2020) 103936.

\bibitem{fan2020defective}
S.-K.~S. Fan, C.-Y. Hsu, C.-H. Jen, K.-L. Chen, L.-T. Juan, Defective wafer
  detection using a denoising autoencoder for semiconductor manufacturing
  processes, Advanced Engineering Informatics 46 (2020) 101166.

\bibitem{ronneberger2015u}
O.~Ronneberger, P.~Fischer, T.~Brox, U-net: Convolutional networks for
  biomedical image segmentation, in: International Conference on Medical image
  computing and computer-assisted intervention, Springer, 2015, pp. 234--241.

\bibitem{pandey2019new}
A.~Pandey, D.~Wang, A new framework for cnn-based speech enhancement in the
  time domain, IEEE/ACM Transactions on Audio, Speech, and Language Processing
  27~(7) (2019) 1179--1188.

\bibitem{zhang2022practical}
K.~Zhang, Y.~Li, J.~Liang, J.~Cao, Y.~Zhang, H.~Tang, R.~Timofte, L.~Van~Gool,
  Practical blind denoising via swin-conv-unet and data synthesis, arXiv
  preprint arXiv:2203.13278.

\bibitem{hao2019multi}
H.~Hao, M.~Liu, P.~Xiong, H.~Du, H.~Zhang, F.~Lin, Z.~Hou, X.~Liu, Multi-lead
  model-based ecg signal denoising by guided filter, Engineering Applications
  of Artificial Intelligence 79 (2019) 34--44.

\bibitem{xiong2016ecg}
P.~Xiong, H.~Wang, M.~Liu, S.~Zhou, Z.~Hou, X.~Liu, Ecg signal enhancement
  based on improved denoising auto-encoder, Engineering Applications of
  Artificial Intelligence 52 (2016) 194--202.

\bibitem{xiong2020novel}
S.~Xiong, H.~Zhou, S.~He, L.~Zhang, Q.~Xia, J.~Xuan, T.~Shi, A novel end-to-end
  fault diagnosis approach for rolling bearings by integrating wavelet packet
  transform into convolutional neural network structures, Sensors 20~(17)
  (2020) 4965.

\bibitem{recoskie2018learning}
D.~Recoskie, R.~Mann, Learning sparse wavelet representations, arXiv preprint
  arXiv:1802.02961.

\bibitem{jawali2019learning}
D.~Jawali, A.~Kumar, C.~S. Seelamantula, A learning approach for wavelet
  design, in: ICASSP 2019-2019 IEEE International Conference on Acoustics,
  Speech and Signal Processing (ICASSP), IEEE, 2019, pp. 5018--5022.

\bibitem{ha2021adaptive}
W.~Ha, C.~Singh, F.~Lanusse, E.~Song, S.~Dang, K.~He, S.~Upadhyayula, B.~Yu,
  Adaptive wavelet distillation from neural networks through interpretations,
  arXiv preprint arXiv:2107.09145.

\bibitem{wang2018multilevel}
J.~Wang, Z.~Wang, J.~Li, J.~Wu, Multilevel wavelet decomposition network for
  interpretable time series analysis, in: Proceedings of the 24th ACM SIGKDD
  International Conference on Knowledge Discovery \& Data Mining, 2018, pp.
  2437--2446.

\bibitem{coifman1992wavelet}
R.~R. Coifman, Y.~Meyer, V.~Wickerhauser, Wavelet analysis and signal
  processing, in: In Wavelets and their applications, Citeseer, 1992.

\bibitem{mallat1999wavelet}
S.~Mallat, A wavelet tour of signal processing, Elsevier, 1999.

\bibitem{strang1996wavelets}
G.~Strang, T.~Nguyen, Wavelets and filter banks, SIAM, 1996.

\bibitem{kingma2014adam}
D.~P. Kingma, J.~Ba, Adam: A method for stochastic optimization, arXiv preprint
  arXiv:1412.6980.

\bibitem{hao2017improved}
H.~Hao, H.~Wang, N.~ur~REHMAN, L.~Chen, H.~Tian, An improved multivariate
  wavelet denoising method using subspace projection, IEICE Transactions on
  Fundamentals of Electronics, Communications and Computer Sciences 100~(3)
  (2017) 769--775.

\bibitem{jimenez2019u}
G.~Jimenez-Perez, A.~Alcaine, O.~Camara, U-net architecture for the automatic
  detection and delineation of the electrocardiogram, in: 2019 Computing in
  Cardiology (CinC), IEEE, 2019, pp. Page--1.

\bibitem{mesaros2018multi}
A.~Mesaros, T.~Heittola, T.~Virtanen, A multi-device dataset for urban acoustic
  scene classification, arXiv preprint arXiv:1807.09840.

\bibitem{fonseca2018general}
E.~Fonseca, M.~Plakal, F.~Font, D.~P. Ellis, X.~Favory, J.~Pons, X.~Serra,
  General-purpose tagging of freesound audio with audioset labels: Task
  description, dataset, and baseline, arXiv preprint arXiv:1807.09902.

\bibitem{coifman1992entropy}
R.~R. Coifman, M.~V. Wickerhauser, Entropy-based algorithms for best basis
  selection, IEEE Transactions on information theory 38~(2) (1992) 713--718.

\end{thebibliography}

\appendix

\section{}
\subsection{Choice of the NN architectures: case function model}\label{AppendixA}
We aim to adapt the AE architecture from \cite{liu2019fault} to our case, the main differences being the number of input samples. The Table~\ref{TableAnn1} shows at the columns "Function", the obtained specialisation score ($S_p$), robustness score ($S_r$) and ($\bar{S} $) mean score for different set of architectural choices. The number of filters for each layers is given in the first column, the kernel size has 5 coefficients in the last layer and increases by 5 for each outer layer. We can remark the robustness score decreases as the number of parameters increase. The architecture [16-32-64-128-256] has the best specialisation score ($S_p$=$22$) compared to [32-64-128-256] with $S_p$=$24$. However, we select the later architecture because it is way more robust ($S_r$=$577$ vs $S_r$=$919$) compared to the small gain in specialisation. 
\begin{table}
\footnotesize
\center
\begin{tabular}{c|ccc|ccc}
\hline
\multirow{2}{*}{Architecture} &    \multicolumn{3}{c|}{Function} &    \multicolumn{3}{c}{Backgorund}   \\ & $S_p$ &  $S_r$ & $\bar{S} $  & $S_p$ &  $S_r$ & $\bar{S} $ \\
\hline 
$\text{[16-32-64]}$&               46& 427& 332 &               18.5& 17.8& 17.9\\
$\text{[32-64-128]}$&             45& 430& 334 &             17.7& 17& 17.2\\
$\text{[64-128-256]}$&           45&  434& 337 &           18.2&  17.9& 18\\
$\text{[16-32-64-128]}$&        25& 590& 448&        17.1& 17.7& 17.7\\
$\text{[32-64-128-256]}$&      24& 577& 439&      16.9& 17.1& 17\\
$\text{[64-128-256-512]}$&    54& 663& 510&    19& 20& 19.9\\
$\text{[16-32-64-128-256]}$&        22& 919& 694&        21.1& 26& 25.4\\
$\text{[32-64-128-256-512]}$&      118& 1058& 823&      21.6& 25.5& 25\\
\hline
\end{tabular} 
\caption{Specialisation ($S_p$), robustness ($S_r$) and mean ($\bar{S} $) score for the Block function ("Function") and for the first fold of the background denoising case ("Background"). }\label{TableAnn1}
\end{table}

\subsection{Parameter summary of the NN architectures}\label{AppendixC}
The Table~\ref{Table3_1} provides an overview of the key parameters of the six different architectures. Unless stated otherwise, the activation function used at the output of each layer is the leaky ReLU with a leak of 0.1. The use of dropout and PRELU proposed in \cite{nossier2020experimental} was tested but abandoned because it decreased the performance. 

\subsection{Function class generation}\label{AppendixD}

The definition of the four function classes is as follows:

$\bullet$ The \textit{Block} class:
\begin{align}
&s(t)=\sum_{i=1}^{N_b} {B(t,\tilde{\tau}_{i},\tilde{w}_i,\tilde{\alpha}_i}) \\
&{\rm with} \hspace{0.2cm} B(t,\tau,w,\alpha) = \begin{cases} \alpha & \text{if $t \in [\tau-\frac{w}{2}, \tau+\frac{w}{2}]$,} \\
                      0 & \text{else} \end{cases}
\end{align}
The random parameters $\tilde{\tau}_{i}$ and $\tilde{w}_i$ are respectively the center and the width of $N_b=10$ block cutting the time axis of the signal $s$. Those blocks are generated by selecting randomly $N_b-1$ samples ${t_2,t_3,...,t_{N_b}}$ from 1 to $T$, listing them by increasing value and by fixing $t_1=0$ and $t_{N_b+1}=T$. Then $\tilde{\tau}_{i}=t_i+\tilde{w}_i/2$ and $\tilde{w}_i =t_{i+1}-t_{i} $. The amplitude $\tilde{\alpha}_i$ is a random variable generated from a normal distribution. 

$\bullet$ The \textit{Bumps} class:
\begin{align}
&s(t)=\sum_{i=1}^{N_b} {B(t,\tilde{\tau}_{i},\tilde{w}_i,\tilde{\alpha}_i}) \\
&{\rm with} \hspace{0.2cm} B(t,\tau,w,\alpha) = \frac{|\alpha |}{(1+\frac{5}{w}|t-\tau|)^4}
\end{align}
The random variables and parameters are the same as in the \textit{Block} class.

$\bullet$ The \textit{HeaviSine} class:
\begin{align}
&s(t)=\sum_{i=1}^{N_b} {B(t,\tilde{\tau}_{i},\tilde{w}_i,\tilde{\alpha}_i,\tilde{f}_i,\tilde{\phi}_i}) \\
&{\rm with} \hspace{0.2cm} B =\begin{cases} |\alpha |{\rm sin}(\frac{ft}{200}+\phi) & \text{if $t \in [\tau-\frac{w}{2}, \tau+\frac{w}{2}]$,} \\
                      0 & \text{else} \end{cases}
\end{align}
For this case, we fix $N_b=4$, the frequency variables $\tilde{f}_i$ and the phase variable $\tilde{\phi}_i$ are both different realisations of a normal distribution. 

$\bullet$ The \textit{Doppler} class:
\begin{align}
s(t)= {\rm pad}_{\tilde{t}_p}\left[ [t(t-1)]^{\frac{1}{\tilde{z}}} {\rm sin}\left( 16\pi \frac{1+0.2}{20t+0.2} \right) \right]
\end{align}
where pad is a zero-padding function adding $\tilde{t}_p$ 0 at the beginning of the signal, then inverting the function half of the realisations and cropping it, so it contains exactly $T$ samples. The padding variable $\tilde{t}_p$ is generated by selecting a random number from 0 to $T/2$, and the power variable $\tilde{z}$ is generated by selecting a random number from 0 to $10$.

In order to keep the signals values in the same range for each class, we normalize each realisation between 0 and 1 by performing the following transformation: 
\begin{align}
s(t)  \leftarrow  \frac{s(t)-s_{\rm min}}{s_{\rm max}-s_{\rm min}}
\end{align}
with $s_{\rm min}$ and $s_{\rm max}$ respectively the minimum and the maximum value of the current realisation $s$.

\section{}
\subsection{Choice of the NN architectures: airport background removal}\label{AppendixB}
The Table~\ref{TableAnn1} column "Background", shows the experiment from Appendix~\ref{AppendixA} applied to the first fold of the Background denoising task. 
We choose the architecture $[32-64-128-256]$ because it provides the best $S_p$ score and close to the best $S_r$ score.

We introduced AE-small as an AE architecture with similar number of parameters than L-WPT. However, the L-WPT with $L=8$ contains more parameters ($n_p=8670$) than when $L=5$. 
For simplicity, we keep using then the small architectures from Table~\ref{Table3_1}. Indeed,
we also tried three architectures with number of parameters around $n_p=8670$: [16-32] with $n_p=5488$, [8-16-32] with $n_p=6632$ and [32-64] with $n_p=21216$. The best performing is the [32-64] with $S_p=19.3$ and $S_r=19.5$ for the first fold. It outperforms the AE-small ($S_p=21.9$, $S_r=22,4$) but does worse than AE-large anyway ($S_p=16.9$, $S_r=17.1$).

\begin{table}
\center
\begin{tabular}{cccc}
\multicolumn{4}{c}{\textbf{CNN-large $n_p=279649$}}  \\ 
\hline 
\textbf{Layers} &\textbf{ Filters} &\textbf{ Kernels} & \textbf{Comments} \\ 
\hline 
Conv & 32 & 20 & No stride  \\ 

Conv & 64 & 15 & No stride \\ 

Conv & 128 & 10 & No stride \\ 

Conv & 256 & 5 & No stride \\ 
 
Conv & 32 & 8 & No stride  \\ 
\hline 
\multicolumn{4}{c}{\textbf{CNN-small $n_p=736$}} \\ 
\hline 
\textbf{Layers} &\textbf{ Filters} &\textbf{ Kernels} & \textbf{Comments} \\ 
Conv & 8 & 10 & No stride \\ 
Conv & 16 & 5 & No stride\\ 
Conv & 8 & 8 & No stride\\ 
\hline 
\multicolumn{4}{c}{} \\
\multicolumn{4}{c}{\textbf{AE-large $n_p=554954$}} \\ 
\hline 
\textbf{Layers} &\textbf{ Filters} &\textbf{ Kernels} & \textbf{Comments}  \\ 
\hline 
Conv & 32 & 20 & -\\ 

Conv & 64 & 15 &  - \\ 

Conv & 128 & 10 &   -  \\ 

Conv & 256 & 5 &  - \\ 
 
T-Conv & 128 & 5 & -  \\ 
 
T-Conv & 64 & 10 & - \\ 

T-Conv & 32 & 15 & -  \\ 

T-Conv & 1 & 20 & Lin act  \\ 
\hline 
\multicolumn{4}{c}{\textbf{AE-small $n_p=1473$}}  \\ 
\hline 
\textbf{Layers} &\textbf{ Filters} &\textbf{ Kernels} & \textbf{Comments} \\ 
Conv & 8 & 10 & - \\ 
Conv & 16 & 5 & - \\ 
T-Conv & 8 & 5 & - \\ 
T-Conv & 1 & 10 & Lin act\\ 
\hline 
\multicolumn{4}{c}{} \\
\multicolumn{4}{c}{\textbf{U-Net-large $n_p=554954$}} \\ 
\hline 
\textbf{Layers} &\textbf{ Filters} &\textbf{ Kernels} & \textbf{Comments}  \\ 
\hline 
Conv & 32 & 20 & SC  \\ 

Conv & 64 & 15 &  SC  \\ 

Conv & 128 & 10 &   SC  \\ 

Conv & 256 & 5 &  -  \\ 
 
T-Conv & 128 & 5 & -  \\ 
 
T-Conv & 64 & 10 &  -   \\ 

T-Conv & 32 & 15 & -   \\ 

T-Conv & 1 & 20 & Lin act  \\ 
\hline 
\multicolumn{4}{c}{\textbf{U-Net-small $n_p=1473$}} \\ 
\hline 
\textbf{Layers} &\textbf{ Filters} &\textbf{ Kernels} & \textbf{Comments} \\ 
Conv & 8 & 10 & SC \\ 
Conv & 16 & 5 & - \\ 
T-Conv & 8 & 5 & - \\ 
T-Conv & 1 & 10 & Lin act\\ 
\hline 
\end{tabular} 
\caption{The configuration of the six implemented neural network architectures. We used the following abbreviations: convolutional layer (Conv),  transposed convolutional layer (T-Conv), linear activation function (Lin act), stride value is set to 1 (No stride), skip connection of the output to the related T-Conv layer (SC). }\label{Table3_1}
\end{table}

\end{document}